\documentclass[11pt]{article}
\usepackage[ansinew]{inputenc} 
\usepackage{amssymb,amsfonts}
\usepackage{amsthm}
\def\nn{\nonumber}

\def \bc {\begin{center}}
\def \ec {\end{center}}
\def \bi {\begin{itemize}}
\def \ei {\end{itemize}}

\def \ba {\begin{array}}
\def \ea {\end{array}}

\def \bea {\begin{eqnarray}}
\def \eea {\end{eqnarray}}

\def \be {\begin{equation}}
\def \ee {\end{equation}}

\def \cL {{\cal W}}
\def \cG {{\cal G}}
\def\cO {{\mathbb O}}
\def \cas {c}

\def\um {\frac{1}{2}}
\def\W {\cal W}

\def\w#1{{v}_#1}
\def\z#1{{z}_#1}
\def\u#1{{u}_#1}

\newcommand{\bra}[1]{ \left<#1\right| }
\newcommand{\ket}[1]{ \left|#1\right> }
\newcommand{\scprod}[2]{ \left<#1\right.|\left.#2\right> }

\newtheorem{thm}{Theorem}[section]

\newtheorem{defn}[thm]{Definition}
\theoremstyle{remark}

\begin{document}
\begin{center}
{\LARGE {\bf Generalized ${\cal W}_{\infty}$ Higher-Spin Algebras and 
Symbolic Calculus on Flag Manifolds}\footnote{Work partially supported by 
the MCYT and Fundación Séneca under projects BFM 2002-00778 and 
PB/9/FS/02}} 
\end{center}
\bigskip
\bigskip

\centerline{{\sc Manuel Calixto}}

\bigskip

\bc {\it Departamento de Matemática Aplicada y Estad\'\i stica, 
Universidad Politécnica de Cartagena, Paseo Alfonso XIII 56, 30203 
Cartagena, Spain} 

\bigskip

E-mail: Manuel.Calixto@upct.es
\ec

\bigskip

\bigskip
\begin{center}
{\bf Abstract}
\end{center}
\small

\begin{list}{}{\setlength{\leftmargin}{3pc}\setlength{\rightmargin}{3pc}}
\item We study a new class 
of infinite-dimensional Lie algebras $\cL_\infty(N_+,N_-)$ generalizing 
the standard $\W_\infty$ algebra, viewed as a tensor operator algebra of 
$SU(1,1)$ in a group-theoretic framework. Here we interpret 
$\cL_\infty(N_+,N_-)$ either as an infinite continuation of the pseudo-unitary 
symmetry $U(N_+,N_-)$, or as a ``higher-$U(N_+,N_-)$-spin extension" of 
the diffeomorphism algebra diff$(N_+,N_-)$ of the $N=N_++N_-$ torus 
$U(1)^N$. We highlight this higher-spin structure of $\cL_\infty(N_+,N_-)$ by developing the 
representation theory of $U(N_+,N_-)$ (discrete series), calculating 
higher-spin representations, coherent states and deriving K\"ahler 
structures on flag manifolds. They are essential ingredients to define 
operator symbols and to infer a geometric pathway between these 
generalized $\W_\infty$ symmetries and algebras of symbols of 
$U(N_+,N_-)$-tensor operators. Classical limits (Poisson brackets on flag manifolds) and quantum (Moyal) 
deformations are also discussed. As potential applications, we comment on 
the formulation of diffeomorphism-invariant gauge field theories, like 
gauge theories of higher-extended objects, and non-linear sigma models on 
flag manifolds. 
\end{list}
\normalsize





\newpage
\section{Introduction}
The long sought-for unification of all interactions and exact solvability 
of (quantum) field theory and statistics parallels the quest for new 
symmetry principles. Symmetry is an essential resource when facing those 
two fundamental problems, either as a gauge guide principle or as a 
valuable classification tool. The representation theory of 
infinite-dimensional groups and algebras has not progressed very far, 
except for some important achievements in one and two dimensions (mainly 
Virasoro, $\W_\infty$ and Kac-Moody symmetries), and necessary 
breakthroughs in the subject remain to be carried out. The ultimate 
objective of this paper is to create a stepping stone to the development 
of a new class of infinite-dimensional symmetries, with potential useful 
applications in (quantum) field theory. 

The structure of the proposed infinite symmetries resembles the one of the 
so-called $\W$ algebras. In the last decade, a large body of literature 
has been devoted to the study of 
$\W$-algebras, and the subject still continues to be fruitful. These 
algebras were first introduced as higher-conformal-spin 
$s> 2$ extensions \cite{Zamolodchikov} of the Virasoro algebra $(s=2)$ 
through the operator product expansion of the stress-energy tensor and 
primary fields in two-dimensional conformal field theory. 
$\W$-algebras have been widely used in two-dimensional physics, mainly in
condensed matter, integrable models
(Korteweg-de Vries, Toda), phase transitions in two dimensions,
stringy black holes and, at a more fundamental level, as the underlying
gauge symmetry  of two-dimensional gravity models
generalizing the Virasoro gauged symmetry in the light-cone discovered
by Polyakov \cite{Poly} by adding spin $s>2$ currents (see e.g.
\cite{Bergshoeff} and \cite{Shen,Hull} for a review). Only when all ($s\to\infty$) 
conformal spins $s\geq 2$ are considered,
the algebra (denoted by $\W_{\infty}$) is proven to be of Lie type;
moreover, currents of spin $s=1$ can also be included \cite{Pope2}, thus
leading to the Lie algebra ${\cal W}_{1+\infty}$, which plays a
determining role in the classification of all universality classes
of incompressible quantum fluids and the identification of the
quantum numbers of the excitations in the quantum Hall effect
\cite{Capelli}.

The process of elucidating the mathematical structure underlying $\W$ 
algebras has led to various directions. Geometric approaches identify the 
classical ($\hbar\to 0$) limit $w_\infty$ of $\W_\infty$ algebras with 
area-preserving (symplectic) diffeomorphism algebras of two dimensional 
surfaces \cite{Bakas,Witten}. These algebras possess a Poisson structure, 
and it is a current topic of great activity to recover the ``quantum 
commutator" $[\cdot,\cdot]$ from (Moyal-like) deformations of the Poisson 
bracket $\{\cdot,\cdot\}$. There is a group-theoretic structure underlying 
these quantum deformations \cite{Pope}, according to which 
${\W}_{\infty}$ algebras are just particular members of a one-parameter family 
${\W}_{\infty}(c)$ of non-isomorphic \cite{Hoppe2,Bergshoeff2} infinite-dimensional Lie-algebras of
$SU(1,1)$ tensor operators (when ``extended beyond the wedge" \cite{Pope} or 
``analytically continued" \cite{Fradkin2}). The (field-theoretic) 
connection with the theory of higher-spin gauge fields in (1+1)- and 
(2+1)-dimensional anti-de Sitter space AdS 
\cite{Fradkin2,Fradkin,Vasiliev} ---homogeneous spaces of $SO(1,2)\sim 
SU(1,1)$ and $SO(2,2)\sim SU(1,1)\times SU(1,1)$, respectively--- is then 
apparent within this group-theoretical context. Also, the relationship 
between area-preserving diffeomorphisms and 
${\W}_{\infty}$ algebras emerges naturally in this group-theoretic picture; indeed, it is well known that 
coadjoint orbits of any semisimple Lie group like $SU(1,1)\simeq 
SL(2,\mathbb{R})$ (cone and hyperboloid of one and two sheets) naturally 
define a symplectic manifold, and the symplectic structure inherited from 
the group can be used to yield a Poisson bracket, which leads to a 
geometrical approach to quantization. From an algebraic point of view, the 
Poisson bracket is the classical limit of the quantum commutator of 
``covariant symbols" (see next Section). However, the essence of the full 
quantum algebra is captured in a classical construction by extending the 
Poisson bracket to Moyal-like brackets. In particular, one can reformulate 
the (cumbersome) problem of calculating commutators of tensor operators of 
$su(1,1)$ in terms of (easier to perform) Moyal (deformed) 
brackets of polynomial functions on coadjoint orbits 
$\cO$ of $SU(1,1)$.  A further simplification, that we shall use, then consists of 
taking advantage of the standard oscillator realization (\ref{bosoprea}) 
of the semisimple Lie algebra generators and replacing non-canonical 
(\ref{poissonlie}) by Heisenberg brackets (\ref{nPoissonbraket}). 

Going from three-dimensional algebras $su(2)$ and $su(1,1)$ to 
higher-dimensional pseudo-unitary algebras $su(N_+,N_-)$ entails 
non-trivial problems. Actually, the classification and labelling of tensor 
operators of Lie groups other than 
$SU(1,1)$ and $SU(2)$ is not an easy task in general. In the letter \cite{infdimal}, the 
author put forth an infinite set ${\cL}_{\infty}(N_+,N_-)$ of tensor 
operators of $U(N_+,N_-)$ and calculated the structure constants of this 
quantum associative operator algebra by taking advantage of the oscillator 
realization of the $U(N_+,N_-)$ Lie-algebra, in terms of 
$N=N_++N_-$ boson operators [see Eq. (\ref{bosoprea})], and by using Moyal brackets. Operator labelling 
coincides here with the standard Gel'fand-Weyl pattern for vectors in the 
carrier space of unirreps of $U(N)$ (see later on Sec. 
\ref{secoherente2}). Later on, the particular case of 
${\cL}_\infty(2,2)$ was identified in \cite{vp} with a four-dimensional 
analogue of the Virasoro algebra, i.e. an infinite extension (``promotion 
or analytic continuation" in the sense of \cite{Fradkin2}) of the 
finite-dimensional conformal symmetry $SU(2,2)\sim SO(4,2)$ in 3+1D. Also, 
${\cL}_\infty(2,2)$ was interpreted as a higher-conformal-spin extension 
of the diffeomorphism algebra diff$(4)$ of vector fields on a 
4-dimensional manifold (just as ${\W}_\infty$ is a higher-spin extension 
of the Virasoro diff$(1)$ algebra), thus constituting a potential gauge 
guide principle towards the formulation of of induced conformal gravities 
(Wess-Zumino-Witten-like models) in realistic dimensions \cite{qg}. For 
completeness, let as say that  
${\W}_\infty$-algebras also appear as central extensions of the algebra of 
(pseudo-)differential operators on the circle \cite{Bakas2}, and 
higher-dimensional analogues have been constructed in that context 
\cite{Ramosh}; however, we do not find a clear connection with our 
construction.

In this article the aim is to infer a concrete pathway between these 
natural (algebraic) generalizations ${\cL}_{\infty}(N_+,N_-)$ of 
${\W}_\infty$, and infinite higher-spin algebras 
of $U(N_+,N_-)$ operator symbols, using the coherent-state machinery and 
tools of geometric and Berezin quantization. In order to justify the view 
of ${\W}_\infty(N_+,N_-)$ as a ``higher-spin algebra'' of $U(N_+,N_-)$, we 
shall develop the representation theory of 
$U(N_+,N_-)$, calculating arbitrary-spin coherent states and deriving K\"ahler structures on flag manifolds, 
which are essential ingredients to define operator symbols, star-products 
and to compute the leading order ($\hbar\to 0$, or large quantum numbers) 
structure constants of star-commutators in terms of Poisson brackets on 
the flag space. Actually, the structure constants calculated in 
\cite{infdimal} were restricted to a class of irreducible representations 
given by oscillator representations. Here we show how to deal with the 
general case.

 Throughout the paper, we shall discuss either classical 
limits of quantum structures (Poisson brackets from star-commutators) or 
quantum deformations of classical objects (Moyal deformations of 
oscillator algebras). 

We believe this paper touches a wide range of different algebraic and 
geometric structures of importance in Physics and Mathematics. Our main 
objective is to describe them and to propose interconnections between 
them. Therefore, except for Section \ref{backdrop}, which summarizes some 
basic definitions and theorems found in the literature, we have rather 
preferred to follow a fairly descriptive approach throughout the paper. 
Perhaps pure Mathematicians will miss the ``Theorem-Proof'' procedure to 
present some of the particular results of this work, but I hope our plan 
will make the presentation more dynamic and will result in greater 
dissemination of the underlying ideas and methods. 

The organization of the paper is as follows. Firstly we set the general 
context of our problem and remind some basic theorems and notions on the 
representation theory of Lie groups (in particular, we focus on 
pseudo-unitary groups) and geometric structures derived from it. In Sec. 
\ref{su2su11} we exemplify the previous structural information with the 
case of three-dimensional underlying algebras $su(2)$ and 
$su(1,1)$, their tensor operator algebras, classical limits, Lie-Poisson structures 
and their relevance in large-$N$ matrix models (and relativistic 
membranes) and ${\cal W}_{(1+)\infty}$ invariant theories. In Sec. 
\ref{ghsa} we extend these constructions to general pseudo-unitary groups 
and we show how to build ``generalized $w_\infty$ algebras" 
$w_\infty(N_+,N_-)$ and to compute their quantum (Moyal) deformations ${\cL}_\infty(N_+,N_-)$ through 
oscillator realizations of the $u(N_+,N_-)$ Lie algebra. Then, in Sec. 
\ref{secoherente} we introduce a local complex parametrization of the 
coset representatives 
$SU(N)/U(1)^N=\mathbb F_{N-1}$ (flag space), we construct 
coherent states and derive K\"ahler structures on flag manifolds. They are 
essential ingredients to discuss symbolic calculus on flag manifolds, and 
to highlight the higher-spin structure of the algebra 
${\cL}_\infty(N_+,N_-)$. In Section \ref{fieldflag} we make some comments 
on the potential role of these infinite-dimensional algebras as residual 
gauge symmetries of extended objects (``$N(N-1)$-branes 
$\mathbb F_{N-1}$") in the light-cone gauge, and formulate non-linear 
sigma models on flag manifolds. Last Section is devoted to conclusions and 
outlook.

\section{The group-theoretical backdrop\label{backdrop}}

Let us start by fixing notation and reminding some definitions and results 
on group, tensor operator, Poisson-Lie algebras, coherent states and 
symbols of a Lie group 
$G$; in particular, we shall focus on pseudo-unitary groups: \be 
G=U(N_+,N_-)=\{g\in M_{N\times N}(\mathbb{C})\,\,/\,\,g\Lambda 
g^\dag=\Lambda\},\,\,\,N=N_++N_-,\ee that is, groups of complex $N\times 
N$ matrices $g$ that leave invariant the indefinite metric 
$\Lambda={\rm diag}(1,\stackrel{N_+}{\dots},1,-1,\stackrel{N_-}{\dots},-1)$.
The Lie-algebra $\cG$ is generated by the step operators
$\hat{X}_{\alpha}^{\beta}$,
\be \cG=u(N_+,N_-)=\{\hat{X}_\alpha^\beta,\;\;{\rm with}\;\;
(\hat{X}_{\alpha}^{\beta})_{\mu}^{\nu}\equiv\hbar
\delta_{\alpha}^\nu\delta^{\beta}_\mu\},\label{pun}\ee (we introduce the 
Planck constant $\hbar$ for convenience) with commutation relations: \be 
\left[{\hat{X}}_{\alpha_1}^{\beta_1},{\hat{X}}_{\alpha_2}^{\beta_2}\right]=\hbar 
(\delta_{\alpha_2}^{\beta_1}{\hat{X}}_{\alpha_1}^{\beta_2}- 
\delta_{\alpha_1}^{\beta_2}{\hat{X}}_{\alpha_2}^{\beta_1}).\label{sunmcom}\ee 

There is a standard \emph{oscillator realization} of these step operators in
terms of $N$ boson operator variables ($\hat{a}^\dag_\alpha,\hat{a}^\beta$), given by:
 \be \hat{X}_{\alpha}^{\beta}=\hat{a}^\dag_\alpha \hat{a}^\beta,\,\,\,[\hat{a}^\beta,\hat{a}^\dag_\alpha]=
\hbar \delta_{\alpha}^{\beta}\mathbb{I},\,\,\,\,\alpha,\beta=1,\dots 
N,\label{bosoprea}\ee which reproduces (\ref{sunmcom}) (we use the metric 
$\Lambda$ to raise and lower indices). Thus, for unitary irreducible 
representations of $U(N_+,N_-)$ we have the conjugation relation: \be 
(\hat{X}_\alpha^\beta)^\dag=\Lambda^{\beta\mu}\hat{X}_\mu^\nu\Lambda_{\nu\alpha}.\label{conjrel}\ee 
(sum over doubly occurring indices is understood unless otherwise stated). 
Sometimes it will be more convenient to use the generators 
$\hat{X}_{\alpha\beta}=\Lambda_{\alpha\mu}\hat{X}^\mu_\beta$ instead of $\hat{X}_\alpha^\beta$, 
for which the conjugation relation (\ref{conjrel}) is simply written as 
$\hat{X}_{\alpha\beta}^\dag=\hat{X}_{\beta\alpha}$, and the commutation relations (\ref{sunmcom}) adopt the form:
\be 
\left[{\hat{X}}_{\alpha_1\beta_1},{\hat{X}}_{\alpha_2\beta_2}\right]=\hbar 
(\Lambda_{\alpha_2\beta_1}{\hat{X}}_{\alpha_1\beta_2}-\Lambda_{\alpha_1\beta_2}{\hat{X}}_{\alpha_2\beta_1}).\label{bosreal}\ee 
The oscillator realization (\ref{bosoprea}) of $u(N_+,N_-)$-generators 
will be suitable for our purposes later on. 

\begin{defn} Let $\cG^\otimes$ be the tensor algebra over $\cG$, and 
${\cal I}$ the ideal of $\cG^\otimes$ generated by $[\hat{X},\hat{Y}]-(\hat{X}\otimes 
\hat{Y}-\hat{Y}\otimes \hat{X})$ where $\hat{X},\hat{Y}\in\cG$. The universal enveloping algebra 
${\cal U}(\cG)$ is the quotient $\cG^\otimes/{\cal I}$.
\end{defn}
\noindent [From now on we shall drop the $\otimes$ symbol in writing 
tensor products] 
\begin{thm} (Poincaré-Birkhoff-Witt) The monomials  
$\hat{X}_{\alpha_1\beta_1}^{k_1}\dots\hat{X}_{\alpha_n\beta_n}^{k_n}$, 
with $k_i\geq 0$, form a basis of ${\cal U}(\cG)$.
\end{thm}
\noindent Casimir operators are especial elements of ${\cal U}(\cG)$, 
which commute with everything. There are 
$N$ Casimir operators for 
$U(N_+,N_-)$, which are written as polynomials of degree $1,2,\dots,N$ of 
step operators as follows: \be 
\hat{C}_1=\hat{X}_{\alpha}^\alpha,\;\;\hat{C}_2=\hat{X}_{\alpha}^\beta 
\hat{X}_{\beta}^\alpha,\;\; \hat{C}_3=\hat{X}_{\alpha}^\beta 
\hat{X}_{\beta}^\gamma \hat{X}_{\gamma}^\alpha, \dots\label{Casimir}\ee 
The universal enveloping algebra ${\cal U}(\cG)$ decomposes into 
\emph{factor or quotient Lie algebras} $\cL_{\cas}(\cG)$ as follows: 
\begin{thm} Let 
\[{\cal I}_{\cas}=\prod_{\alpha=1}^N(\hat{C}_\alpha-\hbar^\alpha\cas_\alpha)
{\cal U}(\cG)\] 
be the ideal generated by the Casimir operators $\hat{C}_\alpha$. 
The quotient $\cL_{\cas}(\cG)={\cal U}(\cG)/{\cal I}_{\cas}$ is a Lie algebra (roughly speaking,
this quotient means that we replace $\hat{C}_\alpha$ by the complex c-number
$C_\alpha\equiv\hbar^\alpha\cas_\alpha$ whenever it appears in the commutators of elements of
${\cal U}(\cG)$). We shall refer to $\cL_{\cas}(\cG)$ as a $c$-tensor operator algebra.
\end{thm}

According to Burnside's theorem \cite{Kirillov}, for some critical values 
$\cas_\alpha=\cas^{(0)}_\alpha$, the infinite-dimensional Lie algebra 
$\cL_{\cas}(\cG)$ ``collapses" to a finite-dimensional one. In a more 
formal language: 

\begin{thm}\label{Burnside}(Burnside) When $\cas_\alpha, \alpha=1,\dots,N$ coincide with the
eigenvalues of $\hat{C}_\alpha$ in a $d_{\cas}$-dimensional irrep $D_{\cas}$ of
$G$, there exists an ideal $\chi\subset\cL_{\cas}(\cG)$ such that
$\cL_{\cas}(\cG)/\chi=sl(d_{\cas},\mathbb{C})$,
or $su(d_{\cas})$, by taking a compact real form of the complex Lie
algebra.
\end{thm}

Another interesting structure related to the previous one is the \emph{group
$C^*$-algebra} $\mathbb C^*(G)$ [in order to avoid some technical
difficulties, let us restrict ourselves to the compact $G$ case in the next 
discussion]:
\begin{defn} Let $C^\infty(G)$ be the set of analytic complex functions $\Psi$ on $G$ ,
\be C^\infty(G)=\left\{\Psi:G\to\mathbb{C},\,\,\, g\mapsto 
\Psi(g)\right\}.\label{analcomp}\ee The group algebra $\mathbb C^*(G)$ is 
a $C^*$-algebra with an invariant associative *-product (convolution 
product): \be (\Psi *\Psi')(g')\equiv\int_G d^Lg\,\Psi(g) 
\Psi'(g^{-1}\bullet g'),\label{convoprod}\ee ($g\bullet g'$ denotes the 
composition group law and  
$d^Lg$ stands for the left Haar measure) and an involution 
$\Psi^*(g)\equiv\bar{\Psi}(g^{-1 })$.
\end{defn}
The conjugate space $R(G)$ of $C^\infty(G)$ consists of all generalized 
functions with compact supports. The space $M_0(G)$ of all regular Borel 
measures with compact support is a subspace of $R(G)$. The set $R(G,H)$ of 
all generalized functions on $G$ with compact supports contained in a 
subgroup $H$ also forms a subspace of $R(G)$. The following theorem (see 
\cite{Kirillov}) reveals a connection between $R(G,\{e\})$ [$e\in G$ 
denotes the identity element] and the enveloping algebra: 
\begin{thm}\label{Schwartz}(L. Schwartz)  The algebra $R(G,\{e\})$ is 
isomorphic to the enveloping algebra ${\cal U}(\cG)$.
\end{thm}
This isomorphism is apparent when we realize the Lie algebra $\cG$ by left 
invariant vector fields $\hat{X}^L$ on $G$ and consider the mapping 
$\Phi:\cG\to R(G), \hat{X}\mapsto\Phi_{\hat{X}}$, defined by the formula 
\be
\langle\Phi_{\hat{X}}|\Psi\rangle\equiv(\hat{X}^L\Psi)(e),\;\;\forall\Psi\in 
C^\infty(G), \ee where $\langle\Phi|\Psi\rangle\equiv\int_G 
d^Lg\,\bar{\Phi}(g) \Psi(g)$ denotes a scalar product and 
$(\hat{X}^L\Psi)(e)$ means the action of $\hat{X}^L$ on $\Psi$ restricted 
to the identity element $e\in G$. One can also verify the relation \be 
\langle\Phi_{\hat{X}_1}*\dots*\Phi_{\hat{X}_n}|\Psi\rangle=(\hat{X}^L_1\dots 
\hat{X}^L_n\Psi)(e), \,\,\forall \Psi\in C^\infty(G),\ee between star 
products in $R(G)$ and tensor products in ${\cal U}(\cG)$: 

Let us comment now on the geometric counterpart of the previous algebraic 
structures, by using the language of geometric quantization. 

The classical limit of the convolution commutator 
$[\Psi,\Psi']=\Psi*\Psi'-\Psi'*\Psi$ corresponds to the Poisson-Lie 
bracket \be \{\psi,\psi'\}_{PL}(g)= \lim_{\hbar\to 
0}\frac{i}{\hbar^2}[\Psi,\Psi'](g)= 
i(\Lambda_{\alpha_2\beta_1}{x}_{\alpha_1\beta_2}-\Lambda_{\alpha_1\beta_2} 
{x}_{\alpha_2\beta_1}) \frac{\partial \psi}{\partial x_{\alpha_1\beta_1}} 
\frac{\partial \psi'}{\partial x_{\alpha_2\beta_2}} \label{poissonlie}\ee 
between smooth functions $\psi\in C^\infty(\cG^*)$ on the coalgebra 
$\cG^*$, where 
$x_{\alpha\beta}, \alpha,\beta=1,\dots,N$  denote a coordinate system in the coalgebra
${\cal G}^*=u(N_+,N_-)^*\simeq \mathbb{R}^{N^2}$, seen as a $N^2$-dimensional vector space.
The ``quantization map" relating $\Psi$ and $\psi$ is symbolically given by the
expression:
\be \Psi(g)=\int_{\cG^*}\frac{d^{N^2}\Theta}{(2\pi\hbar)^{N^2}}
e^{\frac{i}{\hbar}\Theta(\hat{X})}\psi(\Theta),\ee
where $g=\exp(\hat{X})=\exp(x^{\alpha\beta}\hat{X}_{\alpha\beta})$ is an element of
$G$ and $\Theta=\theta_{\alpha\beta}\Theta^{\alpha\beta}$ is an element of
$\cG^*$.

The constraints $\hat{C}_\alpha(x)=C_\alpha=\hbar^\alpha\cas_\alpha$ defined by the
Casimir operators (\ref{Casimir}) (written in terms of the coordinates
$x_{\alpha\beta}$ instead of $\hat{X}_{\alpha\beta}$) induce a foliation
\be\cG^*\simeq \bigcup_{C} \cO_{C}\ee of the coalgebra $\cG^*$ into leaves 
$\cO_C$: coadjoint orbits, algebraic (flag) manifolds (see later on Sec. \ref{secoherente}). 
This foliation is 
the (classical) analogue of the (quantum) standard Peter-Weyl 
decomposition (see \cite{Landsman}) of the group algebra $\mathbb C^*(G)$: 
\begin{thm}(Peter-Weyl) Let $G$ be a compact Lie group. The group algebra $\mathbb C^*(G)$ 
decomposes, \be \mathbb C^*(G)\simeq 
\bigoplus_{\cas\in\hat{G}}\cL_{\cas}({\cal G}),\label{P-W}\ee into factor 
algebras $\cL_{\cas}({\cal G})$, where 
$\hat{G}$ denotes the space of all (equivalence classes of) irreducible 
representations of $G$ of dimension $d_{\cas}$. 
\end{thm}

The  leaves $\cO_{C}$ admit a symplectic
structure  $(\cO_{C},\Omega_{C})$, where $\Omega_{C}$
denotes a closed 2-form (a
K\"ahler form), which can be obtained from a K\"ahler potential
$K_{C}$ as:
\be \Omega_{C}(z,\bar{z})=\frac{\partial^2 K_{C}(z,\bar{z})}{\partial z_{\alpha\beta}
\partial \bar{z}_{\sigma\nu}}dz_{\alpha\beta}\wedge d\bar{z}_{\sigma\nu}=
\Omega^{\alpha\beta;\sigma\nu}_{C}(z,\bar{z})dz_{\alpha\beta}\wedge 
d\bar{z}_{\sigma\nu}, \ee where $z_{\alpha\beta},\,\alpha>\beta$ denotes a 
system of complex coordinates in $\cO_{C}$ (see later on Sec. 
\ref{secoherente1}). 

After the foliation of $C^\infty(\cG^*)$ into Poisson algebras $C^\infty(\cO_{C})$,
the Poisson bracket induced on the leaves $\cO_{C}$ becomes:
\be
\left\{\psi_l^{\cas},\psi_m^{\cas}\right\}_P(z,\bar{z})=\sum_{\alpha_j>\beta_j}
\Omega_{\alpha_1\beta_1;\alpha_2\beta_2}^{C}(z,\bar{z})
\frac{\partial\psi_l^{\cas}(z,\bar{z})}{\partial z_{\alpha_1\beta_1}}
\frac{\partial\psi_m^{\cas}(z,\bar{z})}{\partial 
\bar{z}_{\alpha_2\beta_2}}= 
\sum_{n}f_{lm}^n(\cas)\psi_n^{\cas}(z,\bar{z}),\label{poiscoad} \ee The 
structure constants for (\ref{poiscoad}) can be obtained through the 
scalar product $f_{lm}^{n}(c)= \langle 
{\psi_n^{\cas}}|\{\psi_l^{\cas},\psi_m^{\cas}\}_P\rangle$, with 
integration measure (\ref{intmeasleaf}), when the set 
$\{{\psi}_n^{\cas}\}$ is chosen to be orthonormal. 

To each function $\psi\in
C^\infty(\cO_{C})$, one can assign its Hamiltonian vector field
${H}_\psi\equiv\{\psi,\cdot\}_P$, which is divergence-free and
preserves de natural volume form 
\be d\mu_{C}(z,\bar{z})=(-1)^{\left(\ba{c} n\\ 2\ea\right)}\frac{1}{n!}
\Omega_{C}^n(z,\bar{z}),\;\;2n={\rm dim}(\cO_C).\label{intmeasleaf}\ee

In general, any vector field $H$ obeying $L_H\Omega=0$ (with $L_H\equiv 
i_H\circ d+d\circ i_H$ the Lie derivative) is called locally Hamiltonian. 
The space ${\rm LHam}(\cO)$ of locally Hamiltonian vector fields is a 
subalgebra of the algebra ${\rm sdiff}(\cO)$ of symplectic 
(volume-preserving) diffeomorphisms of $\cO$, and the space ${\rm 
Ham}(\cO)$ of Hamiltonian vector fields is an ideal of ${\rm LHam}(\cO)$. 
The two-dimensional case ${\rm dim}(\cO)=2$ is special because 
${\rm sdiff}(\cO)={\rm LHam}(\cO)$, and the quotient ${\rm LHam}(\cO)/{\rm Ham}(\cO)$ can 
be identified with the first de-Rham cohomology class $H^1(\cO,\mathbb{R})$ of $\cO$ via $H\mapsto i_H\Omega$.

Poisson and symplectic diffeomorphism algebras of $\cO_{C_+}=S^2$ and
$\cO_{C_-}=S^{1,1}$ (the sphere and the
hyperboloid)  appear as the classical limit [small $\hbar$ and large (conformal-)spin
$\cas_\pm=s(s\pm 1)$, so that the curvature radius $C_\pm=\hbar^2c_\pm$ remains finite]:

\begin{eqnarray}\lim_{\stackrel{\cas_+\to\infty}{\hbar\to 0}}{\cL}_{\cas_+}(su(2))&\simeq
&C^\infty (S^2) \simeq {\rm sdiff}(S^2)\simeq su(\infty),\label{largeN}\\ 
\lim_{\stackrel{\cas_-\to\infty}{\hbar\to 0}}{\cL
}_{\cas_-}(su(1,1))&\simeq& C^\infty (S^{1,1}) \simeq {\rm 
sdiff}(S^{1,1})\simeq 
 su(\infty,\infty)\nn\end{eqnarray}
of factor algebras of $SU(2)$  and $SU(1,1)$, respectively 
(see \cite{Fradkin2,Bergshoeff2}).\footnote{The approximation ${\rm sdiff}(S^2)\simeq su(\infty)$ 
is still not well understood and additional work should be done towards 
its satisfactory formulation. In \cite{glinfty} the approach to approximate ${\rm sdiff}(S^2)$ and 
${\rm sdiff}(T^2)$ by $\lim_{N\to\infty}su(N)$ was studied and a weak uniqueness theorem was proved; however, 
whether choices of sets of basis functions on spaces with different topologies do in fact correspond to distinct 
algebras deserves more careful study.}

Let us clarify the classical limits (\ref{largeN}) by making use of the 
{\it operator (covariant) symbols} \cite{Berezin}: \be 
L^{\cas}(z,\bar{z})\equiv\langle{\cas} z|\hat{L}|{\cas} z\rangle, \,\,\, 
\hat{L}\in\cL_{\cas}(\cG), \label{covsymb} \ee constructed as the mean 
value of an operator $\hat{L}\in\cL_{\cas}(\cG)$ in the coherent state 
$|\cas z\rangle$ (see 
later on Sec. \ref{secoherente2} for more details). 
Using the resolution of unity: 
\be \int_{\cO_C} |\cas u\rangle\langle \cas u| d\mu_{C}(u,\bar{u})=1\ee 
for coherent states, one can define the so called \emph{star 
multiplication of symbols} 
$L^\cas_1\star L^\cas_2$ as the symbol of the product $\hat{L}_1\hat{L}_2$ of two 
operators $\hat{L}_1$ and $\hat{L}_2$:
\be
(L^\cas_1\star L^\cas_2)(z,\bar{z})\equiv\langle{\cas} z|\hat{L}_1\hat{L}_2|{\cas} z\rangle=
\int_{\cO_C} L^\cas_1(z,\bar{u}) L^\cas_2(u,\bar{z})
e^{-s^2_c(z,u)}d\mu_{\cas}(u,\bar{u}),\label{symbolprod}\ee
where we introduce the non-diagonal symbols
\be
L^\cas(z,\bar{u})= \frac{\langle{\cas} z|\hat{L}|\cas 
u\rangle}{\langle\cas z|\cas u\rangle}\label{nondigsymb}\ee and 
$s^2_{\cas}(z,u)\equiv -\ln|\langle \cas z|\cas u\rangle|^2$ can be 
interpreted as the square of the distance between the points $z,u$ on the 
coadjoint orbit $\cO_{C}$. Using general properties of coherent states 
\cite{Klauder}, it can be easily seen that $s^2_{\cas}(z,u)\geq 0$ tends 
to infinity with 
$\cas\to\infty$, if $z\not=u$, and equals zero if $z=u$. Thus, one can conclude that, 
in that limit, the domain 
$u\approx z$ gives only a contribution to the integral (\ref{symbolprod}). 
Decomposing the integrand near the point $u=z$ and going to the 
integration over $w=u-z$, it can be seen that the Poisson bracket 
(\ref{poiscoad}) provides the first order approximation to the star 
commutator for large quantum numbers $c$ (small $\hbar$); that is: 
\be
L^\cas_1\star L^\cas_2-L^\cas_2\star L^\cas_1=
i\left\{L^\cas_1,L^\cas_2\right\}_P+{\rm O}(1/c_\alpha),\label{comsymb}\ee 
i.e. the quantities $1/c_\alpha\sim\hbar^\alpha$ (inverse Casimir 
eigenvalues) play the role of the Planck constant $\hbar$, and one uses  
that $ds^2_c=\Omega^{\alpha\beta;\sigma\nu}_{C}dz_{\alpha\beta} 
d\bar{z}_{\sigma\nu}$ (Hermitian Riemannian metric on $\cO_C$). We address 
the reader to Sec. \ref{secoherente} for more details. 

Before going to the general $SU(N_+,N_-)$ case, let us discuss the two 
well known examples of $SU(2)$ and $SU(1,1)$. 
\section{Tensor operator algebras of $SU(2)$ and $SU(1,1)$\label{su2su11}}
\subsection{Tensor operator algebras of $SU(2)$ and large-$N$ matrix
models\label{su2largeN}}

Let  $\hat{J}^{(N)}_i,\,\,i=1,2,3$ be three
$N\times N$ hermitian  matrices
with commutation relations: \be 
\left[\hat{J}^{(N)}_{i},\hat{J}^{(N)}_{j}\right]=i 
\hbar\epsilon_{ijk}\hat{J}^{(N)}_{k},\ee that is, a $N$-dimensional 
irreducible representation of the angular momentum algebra $su(2)$. The 
Casimir operator $\hat{C}_2= 
(\hat{{J}}^{(N)})^2=\hbar^2\frac{N^2-1}{4}\mathbb I_{N\times N}$ is a 
multiple of the $N\times N$ identity matrix $\mathbb I$. The factor 
algebra $\cL_{N}(su(2))$ is generated by the 
$SU(2)$-tensor operators:

\be \hat{T}^I_m(N)\equiv \sum_{\stackrel{i_k=1,2,3}{k=1,\dots,I}}
\kappa^{(m)}_{i_1,\cdots,i_I} \, \hat{J}^{(N)}_{i_1}\cdots \hat{J}^{(N)}_{i_I},
\label{su2to}
\ee
where the upper index $I=1,\dots,N-1$ is the spin label,
$m=-I,\dots,I$ is the third component and the complex coefficients
$\kappa^{(m)}_{i_1,\cdots,i_I}$ are  the components of a symmetric and traceless tensor.
According to Burnside's theorem \ref{Burnside}, the factor algebra 
$\cL_{N}(su(2))$ is isomorphic to $su(N)$. Thus, the commutation 
relations: 
\be
\left[\hat{T}^I_m(N),\hat{T}^J_n(N)\right]=f^{IJl}_{mnK}(N)\hat{T}^K_l(N) \label{su2tocr}\ee
are those of the $su(N)$ Lie algebra, where $f^{IJl}_{mnK}(N)$ symbolize the
structure constants which, for the Racah-Wigner basis of tensor operators
\cite{Biedenharn}, can be written in terms of
Clebsch-Gordan and (generalized) $6j$-symbols \cite{Hoppe,Pope,Fradkin2}.

The formal limit $N\to\infty$ of the commutation relations (\ref{su2tocr}) coincides
with the Poisson bracket
\be \left\{Y^I_m,Y^J_n\right\}_P=
\frac{i}{\sin\vartheta}
\left(\frac{\partial Y^I_m}{\partial\vartheta}
\frac{\partial Y^J_n}{\partial\varphi}-\frac{\partial Y^I_m}{\partial\varphi}
\frac{\partial Y^J_n}{\partial\vartheta}
\right)=f^{IJl}_{mnK}(\infty)Y^K_l\label{spherepb}\ee
between spherical harmonics 
\be Y^I_m(\vartheta,\varphi)\equiv\sum_{\stackrel{i_k=1,2,3}{k=1,\dots,I}}
\kappa^{(m)}_{i_1,\cdots,i_I} \, x_{i_1}\cdots x_{i_I},\ee
which are defined in a similar way to tensor operators (\ref{su2to}), but replacing
the angular momentum operators $\hat{{J}}^{(N)}$ by the coordinates
${x}=(\cos\varphi\sin\vartheta,\sin\varphi\sin\vartheta,\cos\vartheta)$, i.e. its covariant symbols 
(\ref{covsymb}). Indeed, the
large-$N$ structure constants can be calculated through the scalar product (see \cite{Floratos}):
\begin{eqnarray*}\lim_{N\to\infty} f^{IJl}_{mnK}(N)&=&f^{IJl}_{mnK}(\infty)=
\langle Y^K_l|\{Y^I_m,Y^J_n\}_P\rangle\\ &=&\int_{S^2}\sin\vartheta
d\vartheta d\varphi\,
\bar{Y}^K_l(\vartheta,\varphi)\{Y^I_m,Y^J_n\}_P(\vartheta,\varphi).\end{eqnarray*}

The set of Hamiltonian vector fields $H^I_m\equiv \left\{Y^I_m,\cdot\right\}_P$ close the
algebra ${\rm sdiff}(S^2)$ of area-preserving diffeomorphisms of the
sphere, which can be identified with $su(\infty)$ in the (``weak convergence") sense of
\cite{glinfty} --see Eq. (\ref{largeN}).
This fact was used in \cite{Hoppe} to approximate the residual gauge symmetry
${\rm sdiff}(S^2)$ of the relativistic spherical membrane by $su(N)|_{N\to\infty}$.
There is an intriguing connection between this theory and the quantum 
mechanics of space constant (``vacuum configurations") $SU(N)$ Yang-Mills 
potentials \be A_\mu(x)^i_j=\sum_{a=1}^{N^2-1}A^a_\mu(x) (\hat{T}_a)^i_j, 
\,\,\,\,\hat{T}_a=\hat{T}^I_m(N),\,a=1,\dots,N^2-1\label{sunympot}\ee in 
the limit of ``large number of colours'' (large-$N$). Indeed, the 
low-energy limit of the $SU(\infty)$ Yang-Mills action 
\begin{eqnarray} {\cal S}&=&\int {\rm d}^4x \langle F_{\mu\nu}(x)|
F^{\mu\nu}(x)\rangle,\nn\\
F_{\mu\nu}&=&\partial_\mu A_\nu -\partial_\nu A_\mu+\{A_\mu,A_\nu\}_P,\label{suinftyaction}\\
A_\mu(x;\vartheta,\varphi)&=&\sum_{I,m}A^{Im}_\mu(x)
Y^I_m(\vartheta,\varphi),\nn\end{eqnarray}
described by space-constant $SU(\infty)$ vector potentials
$X_\mu(\tau;\vartheta,\varphi)\equiv A_\mu(\tau,\vec{0};\vartheta,\varphi)$,
turns out to reproduce the dynamics of the relativistic spherical
membrane (see \cite{Floratos}). Moreover, space-time constant $SU(\infty)$ vector potentials
$X_\mu(\vartheta,\varphi)\equiv A_\mu(0;\vartheta,\varphi)$ lead to the
Schild action density for (null) strings \cite{Schild}; the
argument that the internal symmetry space of the $U(\infty)$ pure
Yang-Mills theory must be a functional space, actually the space of
configurations of a string, was pointed out in Ref. \cite{Gervais}. Replacing the Sdiff$(S^2)$-gauge 
invariant theory (\ref{suinftyaction}) by a $SU(N)$-gauge invariant theory with vector potentials (\ref{sunympot}) 
then provides a form of regularization. 

We shall see later in Sec. \ref{generalization} how actions for
relativistic symplectic $p$-branes (higher-dimensional coadjoint orbits) 
can be defined for general (pseudo-)unitary groups in a similar way. 

\subsection{Tensor operator algebras of $SU(1,1)$ and ${\cal W}_{(1+)\infty}$ symmetry}

As already stated in the introduction, $\W$ algebras were first introduced as
higher-conformal-spin $(s> 2)$ extensions \cite{Zamolodchikov} of the Virasoro algebra $(s=2)$ 
through the operator product expansion of the stress-energy
tensor and primary fields in two-dimensional conformal field theory. 
Only when all ($s\to\infty$) conformal spins are considered,
the algebra (denoted by ${\cal W}_{\infty}$) is proven to be of Lie type.

Their classical limit $w$ proves to have a space-time origin as 
(symplectic) diffeomorphism algebras and Poisson algebras of functions on 
symplectic manifolds. For example, $w_{1+\infty}$ is related to the 
algebra of area-preserving diffeomorphisms of the cylinder. Actually, let 
us choose the next set of classical functions of the bosonic (harmonic 
oscillator) variables 
$a(\bar{a})=\frac{1}{\sqrt{2}}(q\pm ip)=\rho e^{\pm i\vartheta}$ (we
are using mass and frequency $m=1=\omega$, for simplicity):
\be\ba{l} L^I_{+|n|}\equiv\um
(a\bar{a})^{I-|n|}a^{2|n|}=\um\rho^{2I}e^{2i|n|\vartheta},\\
L^I_{-|n|}\equiv\um
(a\bar{a})^{I-|n|}\bar{a}^{2|n|}=\um\rho^{2I}e^{-2i|n|\vartheta},\ea\label{auaral}
\ee where $n\in \mathbb Z; I\in \mathbb Z^+$. 
A straightforward calculation from the basic Poisson bracket
$\{a,\bar{a}\}=i$ provides the following formal Poisson algebra:
\be
\{L^I_m,L^J_n\}=i\left( \frac{\partial L^{I}_{m} }{\partial a}
\frac{\partial L^{J}_{n}}{\partial \bar{a}}- \frac{\partial
L^{I}_{m} }{\partial \bar{a}} \frac{\partial L^{J}_{n}}{\partial
a}\right)=i(In-Jm)L^{I+J-1}_{m+n}, \label{auaralcom} \ee of
functions $L$ on a two-dimensional phase space (see
\cite{simplin}). As a distinguished subalgebra of
(\ref{auaralcom}) we have the set: \be su(1,1)=\{L_0\equiv
L^1_0=\um a\bar{a},\, L_+\equiv L^1_1=\um a^2,\, L_-\equiv
L^1_{-1}=\um \bar{a}^2\}, \ee which provides an oscillator realization of 
the $su(1,1)$ Lie algebra generators $L_\pm,L_0$, in terms of a single 
bosonic variable, with commutation relations (\ref{su11cr}). With this notation, the
functions $L^I_{m}$ in (\ref{auaral}) can also be written as:
\be
L^I_{\pm |m|}=2^{I-1}(L_0)^{I-|m|}(L_\pm)^{|m|}.\label{auarop} \ee This 
expression will be generalized for arbitrary $U(N_+,N_-)$ groups in Eq. 
(\ref{auarop2}). 

Following on the analysis of distinguished subalgebras of 
(\ref{auaralcom}), we have the ``wedge" subalgebra \be w_\wedge\equiv\{ 
L^I_m, \,\, I-|m|\geq 0\}\label{wedge}\ee of polynomial functions of the 
$sl(2,\mathbb R)$ generators $L_0,L_\pm$, which can be formally extended 
beyond the wedge 
$I-|m|\geq 0$ by considering functions on the punctured complex plane with $I\geq 
0$ and arbitrary $m$. To the last set belong the (conformal-spin-2)
generators $L_n\equiv L^1_n, \,n\in\mathbb Z$, which close the Virasoro algebra without
central extension,
\be
\{L_m,L_n\}=i(n-m)L_{m+n},
\ee
and the (conformal-spin-1) generators
$\phi_m\equiv L^{0}_m$, which close the non-extended Abelian Kac-Moody
algebra,
\be
\{\phi_m, \phi_n\}=0.
\ee
In general, the higher-$su(1,1)$-spin fields $L^{I}_n$ have
``conformal-spin" $s=I+1$ and ``conformal-dimension" $n$ (the eigenvalue of
$L^1_0$).

$w$-algebras have been used as the underlying gauge symmetry  of
two-dimensional gravity models, and induced actions for these
``$w$-gravities'' have been written (see for example
\cite{Bergshoeff}). They turn out to be constrained
Wess-Zumino-Witten models \cite{Nissimov}, as happens with
standard induced gravity. The quantization procedure {\it deforms}
the classical algebra $w$ to the quantum algebra $\W$ due to the
presence of anomalies ---deformations of Moyal type of Poisson and
symplectic-diffeomorphism algebras caused essentially by normal
order ambiguities (see bellow). Also, generalizing the
$SL(2,\mathbb{R})$ Kac-Moody hidden symmetry of Polyakov's induced
gravity, there are $SL(\infty,\mathbb{R})$ and
$GL(\infty,\mathbb{R})$ Kac-Moody hidden symmetries for
$\W_{\infty}$ and  ${\cal W}_{1+\infty}$ gravities, respectively
\cite{Popehidden}. Moreover, as already mentioned, the symmetry
${\cal W}_{1+\infty}$ appears to be useful in the classification
of universality classes in the fractional quantum Hall effect.

The group-theoretic structure underlying these $\W$ algebras was
elucidated in \cite{Pope}, where  $\W_{\infty}$ and
${\cal W}_{1+\infty}$ appeared to be distinct members ($\cas=0$ and
$\cas=-1/4$ cases, respectively) of the 
one-parameter family
${\W}_\infty(\cas)$ of non-isomorphic \cite{Hoppe2,Bergshoeff2}
infinite-dimensional factor Lie-algebras of the $SU(1,1)$ tensor 
operators: \bea \hat{L}^I_{\pm |m|}&\propto& \underbrace{\left[ 
\hat{L}_{\mp},\left[ \hat{L}_{\mp},\dots \left[ 
\hat{L}_{\mp},\right.\right.\right.}_{I-|m|\,\,{\rm times}} 
\left.\left.\left. (\hat{L}_\pm)^I\right] \dots\right] \right] = ({\rm 
ad}_{\hat{L}_{\mp}})^{I-|m|}(\hat{L}_\pm)^I \label{sl2rtensorop}\\ 
&\sim&\hat{L}_0^{I-|m|}\hat{L}_\pm^{|m|}+{\rm O}(\hbar),\nn \eea when 
extended beyond the wedge $I-m\geq 0$. The generators 
$\hat{L}_+=\hat{X}_{12},\hat{L}_-=\hat{X}_{21},\hat{L}_0=(\hat{X}_{22}-\hat{X}_{11})/2$, 
fulfil the standard $su(1,1)$ Lie-algebra commutation relations:
\be
\left[\hat{L}_\pm,\hat{L}_0\right]=\pm
\hbar\hat{L}_\pm\,,\;\;\;\;\;
\left[\hat{L}_+,\hat{L}_-\right]=2\hbar\hat{L}_0,\label{su11cr}
\ee
and $\hat{C}=(\hat{L}_0)^2-\um(\hat{L}_+\hat{L}_-+\hat{L}_-
\hat{L}_+)$ is the  Casimir operator of $su(1,1)$. The structure
constants for $\cL_\cas(su(1,1))$ can be written in terms of
$sl(2,\mathbb{R})$ Clebsch-Gordan coefficients and generalized (Wigner) $6j$-symbols
\cite{Pope,Fradkin2}, and they have the general  form: \be
\left[\hat{L}^I_m,\hat{L}^J_n\right]_\cas = \sum_{r=0}^\infty
\hbar^{2r+1}f^{IJ}_{mn}(2r;\cas)
\hat{L}^{I+J-(2r+1)}_{n+m}+\hbar^{2I}Q_I(n;\cas)
\delta^{I,J}\delta_{n+m,0}\mathbb{I},\label{infq}\ee where
$\mathbb{I}\sim\hat{L}^0_0$ denotes a central generator and the 
central charges $Q_I(n;\cas)$ provide for the existence of central extensions. For example,
$Q_1(n;\cas)=\frac{c}{12}(n^3-n)$ reproduces the typical central
extension in the Virasoro sector $I=1$, and $Q_I(n;\cas)$ supplies central 
charges to all conformal-spins $s=I+1$. Quantum deformations of the 
polynomial or ``wedge" subalgebra (\ref{wedge}) do not introduce true 
central extensions. The inclusion of central terms in (\ref{infq}) 
requires the formal extension of (\ref{wedge}) beyond the wedge 
$I-|m|\geq 0$ (see \cite{Pope}), that is, 
the consideration of non-polynomial functions (\ref{auarop}) on the Cartan 
generator $L_0$.

Central charges provide the essential ingredient required to construct 
invariant geometric action functionals on coadjoint orbits of the 
corresponding groups. When applied to Virasoro and ${\cal W}$ algebras, 
they lead to Wess-Zumino-Witten models for {\it induced conformal 
gravities in $1+1$ dimensions} (see e.g. Ref. \cite{Nissimov}). Also, 
local and non-local versions of the Toda systems emerge, as integrable 
dynamical systems, from a one-parameter family of (``quantum tori Lie") 
subalgebras of $gl(\infty)$ (see \cite{HoppeDS}). Infinite-dimensional 
analogues of rigid tops are discussed in \cite{HoppeDS} too; some of these 
systems give rise to ``quantized" (magneto) hydrodynamic equations of an 
ideal fluid on a torus.

The leading order (${\rm O}(\hbar), r=0$) structure constants
$f^{IJ}_{mn}(0;\cas)=Jm-In$ in (\ref{infq}) reproduce the
classical structure constants in (\ref{auaralcom}). It is also precisely 
for the specific values of $\cas=0$ and $\cas=-\frac{1}{4}$ ($\W_{\infty}$ and
${\cal W}_{1+\infty}$, respectively) that  
the sequence of higher-order terms on the right-hand side of (\ref{infq}) turns out to be zero whenever 
$I+J-2r\leq 2$ and $I+J-2r\leq 1$, respectively. Therefore,  ${\cal W}_{\infty}$ (resp. ${\cal W}_{1+\infty}$) 
can be consistently truncated to a 
closed algebra containing only those generators $\hat{L}^I_m$ with positive conformal-spins $s=I+1 \geq 2$ 
(resp. $s=I+1 \geq 1$).

The higher-order terms (${\rm O}(\hbar^3), r\geq 1$) can be captured
in a classical construction by extending the Poisson bracket
(\ref{auaralcom}) to the Moyal bracket
\be
\left\{L^{I}_{m},L^{J}_{n}\right\}_{{\rm M}}=
{L}^{I}_{m}\star {L}^{J}_{n}-{L}^{J}_{n}\star {L}^{I}_{m}=\sum_{r=0}^{\infty}
2\frac{(\hbar/2)^{2r+1}}{(2r+1)!}P^{2r+1}(L^{I}_{m},L^{J}_{n})\,,\label{Moyal}
\ee
where $L\star L'\equiv\exp(\frac{\hbar}{2} P)(L,L')$ is an
{\it invariant associative $\star$-product} and
\be
P^r(L,L')\equiv\Upsilon_{\imath_1\jmath_1}\dots\Upsilon_{\imath_r\jmath_r}
\frac{\partial^r L}{\partial x_{\imath_1}\dots\partial
x_{\imath_r}}\frac{\partial^rL'}{\partial x_{\jmath_1}\dots\partial
x_{\jmath_r}}\,,\label{star} \ee with $x\equiv(a,\bar{a})$ and 
$\Upsilon\equiv \left(\begin{array}{cc} 0 & 1 \\ -1 &0\ea\right)$. We set 
$P^0(L,L')\equiv L\cdot L'$, the ordinary (commutative) product of 
functions. Indeed, Moyal brackets where identified in \cite{Fairlie} as 
the primary quantum deformation ${\cal W}_\infty$ of the classical algebra 
$w_\infty$ of area-preserving diffeomorphisms of the cylinder. 
Also, the oscillator realization in (\ref{auaral}) of the 
$su(1,1)$ Lie-algebra generators $L_\pm,L_0$ in terms of a single boson 
$(a,\bar{a})$ is related to the the ``symplecton'' algebra ${\W}_\infty(-{3}/{16})$ of Biedenharn 
and Louck \cite{Biedenharn} and the higher-spin algebra hs(2) of Vasiliev 
\cite{Vasiliev}. 

\section{Extending the previous constructions to $U(N_+,N_-)$\label{ghsa}}
\subsection{Generalized $w_\infty$ algebras}
The generalization of previous constructions to arbitrary unitary groups 
proves to be quite unwieldy, and a canonical classification of 
$U(N)$-tensor operators has, so far, been proven to exist only for $U(2)$ 
and $U(3)$ (see \cite{Biedenharn} and references therein). Tensor 
labelling is provided in these cases by the Gel'fand-Weyl pattern for 
vectors in the carrier space of unitary irreducible representations of 
$U(N)$ (see later on Sec. \ref{secoherente2}). 

In the letter \cite{infdimal}, a set of  
$U(N_+,N_-)$-tensor operators was put forward and the
Lie-algebra structure constants, for the particular case of the oscillator 
realization (\ref{bosoprea}), were calculated through Moyal bracket (see 
later on Sec. \ref{qnPoissonsec}). The chosen set of operators 
$\hat{L}^I_m$ in the universal enveloping algebra ${\cal U}(u(N_+,N_-))$ 
was a natural generalization of the $su(1,1)$-tensor operators of Eq. 
(\ref{auarop}), where now $L_0$ is be replaced by $N$ Cartan generators 
$\hat{X}_{\alpha\alpha}, \alpha=1,\dots,N$, and $L_+$, $L_-$ are replaced by $N(N-1)/2$ ``rising"  
generators $\hat{X}_{\alpha\beta}, \alpha<\beta$ and $N(N-1)/2$ 
``lowering" generators $\hat{X}_{\alpha\beta}, \alpha>\beta$, 
respectively. The explicit form of these operators is: \bea 
\hat{L}^{I}_{+|m|}&\equiv& 
\prod_{\alpha}(\hat{X}_{\alpha\alpha})^{I_\alpha-(\sum_{\beta>\alpha} 
|m_{\alpha\beta}|+\sum_{\beta<\alpha}|m_{\beta\alpha}|)/2} 
\prod_{\alpha<\beta} (\hat{X}_{\alpha\beta})^{|m_{\alpha\beta}|}, \nn\\ 
 \hat{L}^{I}_{-|m|}&\equiv&
\prod_{\alpha}(\hat{X}_{\alpha\alpha})^{I_\alpha-(\sum_{\beta>\alpha}
|m_{\alpha\beta}|+\sum_{\beta<\alpha}|m_{\beta\alpha}|)/{2}}
\prod_{\alpha<\beta}
(\hat{X}_{\beta\alpha})^{|m_{\alpha\beta}|},\label{auarop2}. \eea
The upper (generalized spin) index
$I\equiv(I_1,\dots,I_N)$ of $\hat{L}$ in (\ref{auarop2}) represents now a
$N$-dimensional vector, which is taken to lie on
a half-integral lattice $I_\alpha\in \mathbb N/2$; the lower index 
(``third component") $m$ symbolizes now an integral upper-triangular 
$N\times N$ matrix, 
\be
m=\left(\ba{ccccc} 0 & m_{12}& m_{13} & \dots & m_{1N}\\
0 & 0 & m_{23} & \dots & m_{2N} \\ 0 & 0 & 0 & \dots & m_{3N} \\
 \vdots & \vdots & \vdots & \ddots  & \vdots \\
\vdots & \vdots & \vdots & \ddots & 0 \ea \right)_{N\times N}, 
m_{\alpha\beta}\in \mathbb Z\label{uppert}\ee and $|m|$ means absolute 
value of all its entries. Thus, the operators $\hat{L}^I_m$ are labelled 
by $N+N(N-1)/2=N(N+1)/2$ indices, in the same way as wave functions 
$\psi^I_m$ in the carrier space of unirreps of $U(N)$ (see Sec. 
\ref{secoherente2}). We shall not restrict ourselves to polynomial 
(``wedge") subalgebras \be 
{\cL}_\wedge(N_+,N_-)\equiv\{\hat{L}^I_m,\,\,\,I_\alpha-(\sum_{\beta>\alpha} 
|m_{\alpha\beta}|+\sum_{\beta<\alpha}|m_{\beta\alpha}|)/2\in\mathbb 
N\}\label{wedge2}\ee and we shall consider ``extensions beyond the wedge" 
(\ref{wedge2}) [to use the same nomenclature as the authors of Ref. 
\cite{Pope} in the context of $\W$ algebras]; that is, we shall let the 
upper indices 
$I_\alpha$ take arbitrary half-integer values $I_\alpha\in \mathbb N/2$. This 
way, we are giving the possibility of true central extensions to the Lie 
algebra (\ref{commu}).\footnote{This claim deserves more careful study. So 
far, it is just an extrapolation of what happens to $\W_\infty$, Virasoro 
and Kac-Moody algebras, where Laurent (and not Taylor or polynomial) 
expansions provide couples of conjugated variables (positive and negative 
modes).} 

The manifest expression of the structure constants $f$ for the
commutators
\be
\left[\hat{L}^{I}_{m},\hat{L}^{J}_{n}\right]=
\hat{L}^{I}_{m}\hat{L}^{J}_{n}-\hat{L}^{J}_{n}\hat{L}^{I}_{m}=
f^{IJl}_{mnK}\hat{L}^{K}_{l}\label{commu}
\ee
of a pair of operators (\ref{auarop2}) 
entails a cumbersome and awkward computation, because of inherent
ordering problems.  However,
the essence of the full ``quantum" algebra (\ref{commu}) 
can be still captured in a classical construction by extending
the Poisson-Lie bracket (\ref{poissonlie})
of a pair of functions $L^{I}_{m},L^{J}_{n}$ on the
commuting coordinates $x_{\alpha\beta}$
to its deformed version, in the sense of Ref. \cite{Bayen}.
To perform calculations with (\ref{poissonlie}) is still rather complicated because of non-canonical
brackets for the generating elements $x_{\alpha\beta}$. A way out to this technical problem is
to make use of the classical analogue of the standard oscillator 
realization, $x_{\alpha\beta}= \bar{a}_\alpha {a}_\beta$, of the generators of
$u(N_+,N_-)$, and replace the Poisson-Lie bracket (\ref{poissonlie}) by the standard Poisson bracket 
 \be
\left\{L^{I}_{m},L^{J}_{n}\right\}=i\Lambda_{\alpha\beta}\left(
\frac{\partial L^{I}_{m} }{\partial a_{\alpha}} \frac{\partial 
L^{J}_{n}}{\partial \bar{a}_{\beta}}- \frac{\partial L^{I}_{m} }{\partial 
\bar{a}_{\beta}} \frac{\partial L^{J}_{n}}{\partial 
a_{\alpha}}\right),\label{nPoissonbraket} \ee for the Heisenberg-Weyl 
algebra $\{{a}_\alpha,\bar{a}_\beta\}=i\Lambda_{\alpha,\beta}$. Although 
it is clear that, in general, both algebras are not isomorphic, the 
difference entails a minor ordering problem, as we are going to show now. 
Moreover, the bracket (\ref{nPoissonbraket}) has the advantage that 
simplifies calculations and expressions greatly. Indeed, it is not 
difficult to compute (\ref{nPoissonbraket}) which, after some algebraic 
manipulations, gives: \be \left\{L^{I}_{m},L^{J}_{n}\right\} 
=i\Lambda^{\alpha\beta}(I_\alpha n_\beta-J_\alpha m_\beta) 
L^{I+J-\delta_{\alpha}}_{m+n},\label{nPoisson} \ee where \be 
m_\alpha\equiv(\sum_{\beta>\alpha}m_{\alpha\beta}- 
\sum_{\beta<\alpha}m_{\beta\alpha})\label{vecupper} \ee defines the 
components of a $N$-dimensional integral vector linked to the integral 
upper-triangular matrix $m$ in (\ref{uppert}), and 
\be
\delta_{\alpha}\equiv(\delta_{\alpha}^1,\dots,\delta_{\alpha}^N) 
\label{deltadef} \ee is a $N$-dimensional vector with the $\alpha^{\rm 
th}$ entry equal to one and zero elsewhere. There is a clear resemblance 
between the $w_\infty$ algebra (\ref{auaralcom}) and (\ref{nPoisson}), 
although the last one is far richer, as we shall show in Section 
\ref{distsubal}. We shall refer to (\ref{nPoisson}) as 
$w_{\infty}(N_+,N_-)$, or ``generalized $w_\infty$", algebra. 

Let us see more carefully what we miss by replacing the Poisson-Lie bracket (\ref{poissonlie}) with the standard 
Poisson bracket (\ref{nPoissonbraket}). First we note that the change of variable 
$x_{\alpha\beta}= \bar{a}_\alpha {a}_\beta$ in
\begin{eqnarray}
&&\Lambda_{\alpha\beta}\left(
\frac{\partial L }{\partial a_{\alpha}}
\frac{\partial L'}{\partial \bar{a}_{\beta}}-
\frac{\partial L }{\partial \bar{a}_{\beta}}
\frac{\partial L'}{\partial a_{\alpha}}\right)=\nn\\
&&\Lambda_{\alpha\beta}\left(
\frac{\partial x_{\alpha_1\beta_1} }{\partial a_{\alpha}}
\frac{\partial L }{\partial x_{\alpha_1\beta_1}}
\frac{\partial x_{\alpha_2\beta_2}}{\partial \bar{a}_{\beta}}
\frac{\partial L' }{\partial x_{\alpha_2\beta_2}}-
\frac{\partial x_{\alpha_1\beta_1} }{\partial \bar{a}_{\beta}}
\frac{\partial L }{\partial x_{\alpha_1\beta_1}}
\frac{\partial x_{\alpha_2\beta_2}}{\partial {a}_{\alpha}}
\frac{\partial L' }{\partial x_{\alpha_2\beta_2}}\right)=\nn\\
&&\Lambda_{\alpha\beta}\left(
\bar{a}_{\alpha_1}\delta_{\beta_1}^\alpha
\frac{\partial L }{\partial x_{\alpha_1\beta_1}}
{a}_{\beta_2}\delta_{\alpha_2}^\beta
\frac{\partial L' }{\partial x_{\alpha_2\beta_2}}-
{a}_{\beta_1}\delta_{\alpha_1}^\beta
\frac{\partial L }{\partial x_{\alpha_1\beta_1}}
\bar{a}_{\alpha_2}\delta_{\beta_2}^\alpha
\frac{\partial L' }{\partial x_{\alpha_2\beta_2}}\right)=\nn\\
&&(\Lambda_{\alpha_2\beta_1}{x}_{\alpha_1\beta_2}-\Lambda_{\alpha_1\beta_2}
{x}_{\alpha_2\beta_1})
\frac{\partial L}{\partial x_{\alpha_1\beta_1}}
\frac{\partial L'}{\partial x_{\alpha_2\beta_2}} ,
\end{eqnarray}
is not one-to-one, as we have $N^2$ (real) coordinates $x_{\alpha\beta}$ and $2N$ (real) coordinates 
$a_\alpha,\bar{a}_\beta$. Also, the Poisson algebra (\ref{nPoissonbraket}) does not distinguish between
polynomials like $x_{\alpha_1\beta_1} x_{\alpha_2\beta_2}$ and
$x_{\alpha_1\beta_2}x_{\alpha_2\beta_1}$, which admit the same
form when written in terms of the commuting oscillator variables
$a_\alpha,\bar{a}_\beta$ as $x_{\alpha\beta}= \bar{a}_\alpha {a}_\beta$.  That is, non-zero combinations like 
$x_{\alpha_1\beta_1} x_{\alpha_2\beta_2}-x_{\alpha_1\beta_2}x_{\alpha_2\beta_1}$ 
behave as zero under Poisson brackets (\ref{nPoissonbraket}). Nevertheless, this is a minor ordering problem 
because we can see that ``null-type" polynomials like:
\be
x_{\alpha_1\beta_1\alpha_2\beta_2}\equiv x_{\alpha_1\beta_1}
x_{\alpha_2\beta_2}-x_{\alpha_1\beta_2}x_{\alpha_2\beta_1}\label{reord}\ee
generate ideals of the algebra $C^\infty(\cG^*)$ of smooth functions 
$L$ on the coalgebra $\cG^*$. Indeed, it suffices to realize that the 
Poisson-Lie bracket between a generic monomial $x_{\alpha\beta}$ and a 
null-type polynomial (\ref{reord}) gives a combination of null-type 
polynomials, that is: \bea 
\left\{x_{\alpha\beta},x_{\alpha_1\beta_1\alpha_2\beta_2}\right\}_{PL}&=& 
i\Lambda_{\alpha_1\beta}x_{\alpha\beta_1\alpha_2\beta_2}- 
i\Lambda_{\alpha\beta_1}x_{\alpha_1\beta\alpha_2\beta_2}+\nn\\ & 
&i\Lambda_{\alpha_2\beta}x_{\alpha_1\beta_1\alpha\beta_2}- 
i\Lambda_{\alpha\beta_2}x_{\alpha_1\beta_1\alpha_2\beta},\eea and 
similarly for general null-type polynomials of higher degree. Thus, we can 
say that the standard Poisson algebra 
(\ref{nPoissonbraket},\ref{nPoisson}) is a subalgebra of the quotient 
$C^\infty(\cG^*)/\cal I$ of $C^\infty(\cG^*)$ [with Poisson-Lie bracket (\ref{poissonlie})] by the 
ideal $\cal I$ generated by null-type polynomials. 

This approximation captures the essence of the full algebra and will be 
enough for our purposes. We shall give in Sec. \ref{secoherente} the main 
guidelines to deal with the general case (general representations). 

Before discussing quantum (Moyal) deformations of (\ref{nPoisson}), let us 
recognize some of its relevant subalgebras. 

\subsection{Distinguished subalgebras of $w_{\infty}(N_+,N_-)$\label{distsubal}}

There are many possible ways of embedding the 
$u(N_+,N_-)$ generators (\ref{bosreal}) inside (\ref{nPoisson}), as there are also 
many possible choices of $su(1,1)$ inside (\ref{auaralcom}). However, 
a ``canonical'' choice is: 
\be
\hat{X}_{\alpha\beta}\equiv -i\hbar
L^{\delta_\alpha}_{e_{\alpha\beta}}\,, \;\;\;
e_{\alpha\beta}\equiv {\rm sign}(\beta-\alpha)
\sum_{\sigma=\min(\alpha,\beta)}^{\max(\alpha,\beta)-1} e_{\sigma,\sigma+1},\label{embedding}
\ee
where $\delta_{\alpha}$ is defined in (\ref{deltadef}) and $e_{\sigma,\sigma+1}$ 
denotes an upper-triangular matrix with 
the $(\sigma,\sigma+1)$-entry equal to one and zero elsewhere, that is 
$(e_{\sigma,\sigma+1})_{\mu\nu}=\delta_{\sigma,\mu}\delta_{\sigma+1,\nu}$ 
(we set $e_{\alpha\alpha}\equiv 0$). 
For example, the $u(1,1)$ Lie-algebra generators correspond to:
\be\ba{ll}
\hat{X}_{12}=-i\hbar L^{(1,0)}_{\left(\begin{array}{cc} 0 &1\\ 0&0\ea\right)},& 
\hat{X}_{21}=-i\hbar L^{(0,1)}_{\left(\begin{array}{cc} 0 &-1\\ 0&0\ea\right)},\\
\hat{X}_{11}=-i\hbar L^{(1,0)}_{\left(\begin{array}{cc} 0 &0\\ 0&0\ea\right)},&
\hat{X}_{22}=-i\hbar L^{(0,1)}_{\left(\begin{array}{cc} 0 &0\\ 0&0\ea\right)}.\ea
\ee
Letting the lower-index $m=e_{\alpha\beta}$ in (\ref{embedding}) run  
over arbitrary integral upper-triangular matrices $m$, we arrive to the  
following infinite-dimensional algebra (as can be seen from 
(\ref{nPoisson})):
\be
\left\{L^{\delta_\alpha}_{m}, L^{\delta_\beta}_{n}\right\}=-i(m^\beta 
L^{\delta_\alpha}_{m+n}-n^\alpha L^{\delta_\beta}_{m+n}) 
\,,\label{difeounm} \ee which we shall denote by 
$w_{\infty}^{(1)}(N_+,N_-)$. Reference \cite{Fradkin2} also considered 
infinite continuations of the particular finite-dimensional symmetries 
$SO(1,2)$ and $SO(3,2)$, as an ``analytic continuation", i.e. an extension 
(or ``revocation", to use their own expression) of the region of 
definition of the Lie-algebra generators' labels. It is easy to see that, 
for $u(1,1)$, the ``analytic continuation'' (\ref{difeounm}) leads to two 
Virasoro sectors: $L_{m_{12}}\equiv L^{(1,0)}_m,\, \bar{L}_{m_{12}}\equiv 
L^{(0,1)}_m$. Its $3+1$ dimensional counterpart 
$w_{\infty}^{(1)}(2,2)$ contains four non-commuting Virasoro-like sectors 
$w_{\infty}^{(1_\alpha)}(2,2)=\{L^{\delta_\alpha}_{m}\}
,\,\alpha=1,\dots,4$ which, in their turn, 
hold three genuine Virasoro sectors for $m=k u_{\alpha\beta},\, 
k\in \mathbb Z,\, \alpha<\beta=2,\dots,4$, where $u_{\alpha\beta}$ denotes an 
upper-triangular matrix with components 
$(u_{\alpha\beta})_{\mu\nu}=\delta_{\alpha,\mu}\delta_{\beta,\nu}$. 
In general, $w_{\infty}^{(1)}(N_+,N_-)$ contains $N(N-1)$ distinct and 
non-commuting Virasoro sectors, \be \left\{V_k^{(\alpha\beta)}, 
V_l^{(\alpha\beta)}\right\}=-i \Lambda^{\alpha\alpha}{\rm 
sign}(\beta-\alpha)\, (k-l)V_{k+l}^{(\alpha\beta)}\,, 
\;\;\;\;\;V_k^{(\alpha\beta)}\equiv L^{\delta_\alpha}_{k u_{\alpha\beta}} 
\ee and holds $u(N_+,N_-)$ as the {\it maximal finite-dimensional 
subalgebra}.

The algebra $w_{\infty}^{(1)}(N_+,N_-)$ can be seen as the {\it minimal} 
infinite continuation of $u(N_+,N_-)$ representing the diffeomorphism 
algebra diff$(N)$ of the $N$-torus $U(1)^N$. Indeed, the algebra 
(\ref{difeounm}) formally coincides with the algebra of vector fields 
$L^\mu_{f(y)}=f(y)\frac{\partial}{\partial y_\mu}$, where 
$y=(y_1,\dots,y_N)$ denotes a local system of coordinates and $f(y)$ 
can be expanded in a plane wave basis, such that 
$L^\mu_{\vec{m}}=e^{im^\alpha y_\alpha}
\frac{\partial}{\partial y_\mu}$ 
constitutes a basis of vector fields for  the so called 
generalized Witt algebra \cite{Ree}, 
\be \left[L^\alpha_{\vec{m}},L^\beta_{\vec{n}}\right]=-i(m^\beta L^{\alpha}_{\vec{m}+\vec{n}}-
n^\alpha L^{\beta}_{\vec{m}+\vec{n}}),\label{witt}\ee
of which there are studies about its representations (see e.g. 
\cite{ramostorus}). Note that, for us, the $N$-dimensional 
lattice vector $\vec{m}=(m_1,\dots,m_N)$ in (\ref{vecupper}) is, by 
construction, constrained to 
$\sum_{\alpha=1}^N m_\alpha=0$ (i.e. $L^\mu_{\vec{m}}$ is \emph{divergence free}), which 
introduces some novelties in (\ref{difeounm}) as regards the Witt algebra 
(\ref{witt}). Actually, the algebra (\ref{difeounm}) can be split into one 
``temporal'' piece, constituted by an Abelian ideal generated by 
$\tilde{L}^N_m\equiv \Lambda_{\alpha\alpha} L^{\delta_\alpha}_{m}$, and a 
``residual'' symmetry generated by the spatial diffeomorphisms 
\be
\tilde{L}^j_m\equiv\Lambda_{jj} L^{\delta_j}_{m}-\Lambda_{j+1,j+1} 
L^{\delta_{j+1}}_{m},\,j=1,\dots,N-1\,\, ({\rm no \ sum \ on \ } j)\,, \ee 
which act semi-directly on the temporal part. More precisely, the 
commutation relations (\ref{difeounm}) in this new basis adopt the 
following form: \bea \left\{ \tilde{L}^j_m,\tilde{L}^k_n\right\} 
&=&-i(\tilde{m}^k \tilde{L}^j_{m+n} - \tilde{n}^j 
\tilde{L}^k_{m+n})\,,\nn\\ \left\{ \tilde{L}^j_m,\tilde{L}^N_n\right\} &=& 
i\tilde{n}^j \tilde{L}^N_{m+n}\,, \label{inftempesp}\\ \left\{ 
\tilde{L}^N_m,\tilde{L}^N_n\right\} &=& 0\,,\nn \eea where 
$\tilde{m}_k\equiv m_k-m_{k+1}$. Only for $N=2$, the last commutator 
admits a central extension of the form 
$\sim n_{12}\delta_{m+n,0}$ 
compatible with the rest of commutation relations 
(\ref{inftempesp}). This result amounts to the fact that the 
(unconstrained) diffeomorphism algebra diff$(N)$ does not admit any 
non-trivial central extension except when $N=1$ (see \cite{nocentral}).

Another important point is in order here. The expression (\ref{embedding}) 
reveals an embedding of the Lie algebra 
$u(N_+,N_-)$ inside the diffeomorphism algebra diff$(N_+,N_-)$ with 
commutation relations (\ref{difeounm}). That is, this new way of labelling 
$u(N_+,N_-)$ generators provides an straightforward ``analytic 
continuation" from $u(N_+,N_-)$ to diff$(N_+,N_-)$. 

As well as  the ``$U(N_+,N_-)$-spin $I=\delta_\mu$ currents" (diffeomorphisms) 
$L^{\delta_\mu}_{m}$ in (\ref{difeounm}), one can also introduce 
``higher-$U(N_+,N_-)$-spin $I$ currents" $L^I_m$ (in a sense 
similar to that of Ref. \cite{Fradkin}) by letting the upper-index $I$ run over an arbitrary 
half-integral $N$-dimensional lattice. Diffeomorphisms 
$L^{\delta_\mu}_{m}$ act semi-directly on ``$u(N_+,N_-)$-spin $J$ currents" $L^J_n$ as 
follows (see Eq. (\ref{nPoisson})):
\be
\left\{L^{\delta_\mu}_{m},L^J_n\right\}=-i\Lambda^{\alpha\beta}J_\alpha 
m_\beta L^{J+\delta_\mu-\delta_\alpha}_{m+n}+in^\mu 
L^J_{m+n}.\label{difaction} \ee Note that this action leaves stable 
Casimir quantum numbers like the trace 
$\sum_{\alpha=1}^NJ_\alpha$ [Casimir $C_1$ eigenvalue (\ref{Casimir})]. 
This higher-spin structure of the algebra $w_\infty(N_+,N_-)$ will be 
justified and highlighted in Section \ref{secoherente}, where higher-spin 
representations of pseudo-unitary groups will be explicitly calculated. 

\subsection{Quantum (Moyal) deformations\label{qnPoissonsec}}

As it happens with $w_\infty$-algebras, the quantization procedure, which 
entails unavoidable renormalizations (mainly due to ordering problems), 
must deform the classical ($\hbar\to 0$) ``generalized $w_\infty$" algebra 
$w_\infty(N_+,N_-)$ in (\ref{nPoisson}) to a quantum algebra 
$\cL_\infty(N_+,N_-)$, by adding higher-order (Moyal-type) terms and 
central extensions like in (\ref{infq}). There is basically only one 
possible deformation $\cL_\infty(N_+,N_-)$ of the bracket 
(\ref{nPoissonbraket}) ---corresponding to a full symmetrization--- that 
fulfils the Jacobi identities (see Ref. \cite{Bayen}), which is the Moyal 
bracket (\ref{Moyal},\ref{star}), where now 
\[\Upsilon\equiv
\left(\begin{array}{cc} 0 & \Lambda \\ -\Lambda &0\ea\right)\] is a $2N\times 2N$ symplectic 
matrix. The calculation of higher-order terms in (\ref{Moyal}) is an 
arduous task, but the result can be summed up as follows:
\be
\left\{L^{I}_{m},L^{J}_{n}\right\}_{{\rm M}}=
\sum_{r=0}^{\infty}2(\frac{\hbar}{2})^{2r+1}f^{\alpha_1\dots\alpha_{2r+1}}_{\alpha_1\dots\alpha_{2r+1}}(I,m;J,n)
L^{I+J-\sum_{j=1}^{2r+1} \delta_{\alpha_j}}_{m+n},\label{qnPoisson}
\ee
where the higher-order structure constants
\be
f^{\alpha_1\dots\alpha_{2r+1}}_{\alpha_1\dots\alpha_{2r+1}}(I,m;J,n)\equiv\sum_{\ell=0}^{2r+1}
\frac{(-1)^\ell}{(2r+1-\ell)!\ell!}\prod_{s=1}^{2r+1}\Lambda^{\alpha_s\beta_s}
\Gamma_{\alpha_s}^\ell(I,-m)\Gamma_{\beta_s}^\ell(J,n)\label{hosc} \ee are 
expressed in terms of the factors 
\be
\Gamma_{\alpha_s}^\ell(I,m)\equiv I^{(s)}_{\alpha_s}+
(-1)^{\theta(\ell-s)}m_{\alpha_s}/2,\label{factors} \ee which are defined 
through the vectors (\ref{vecupper}) and $U(N_+,N_-)$-spins 
\be
I^{(s)}_{\alpha_s}= I_{\alpha_s}-\sum_{t=\theta(s-\ell-1)\ell+1}^{s-1} \delta_{\alpha_s}^{\alpha_t}\,,\;\;\;
I^{(0)}=I^{(\ell+1)}\equiv I,
\ee
with
\be
\theta(\ell-s)=\left\{\begin{array}{l} 0\,\;\;{\rm if}\;\;\ell<s
\\ 1\,\;\;{\rm if}\;\;\ell\geq s\ea\right.
\ee
the Heaviside function. For example, for $r=0$, the leading order (classical, $\hbar\to 0$) structure 
constants are:
\bea
f^{\alpha}_{\alpha}(I,m;J,n)&=&\Lambda^{\alpha\beta}(\Gamma_{\alpha}^0(I,-m)\Gamma_{\beta}^0(J,n)-
\Gamma_{\alpha}^1(I,-m)\Gamma_{\beta}^1(J,n))\nn\\
&=&\Lambda^{\alpha\beta}((I_{\alpha}-m_\alpha/2)(J_{\beta}+n_\beta/2)-
(I_{\alpha}+m_\alpha/2)(J_{\beta}-n_\beta/2)),
\label{leading}
\eea
which, after simplification, coincides with (\ref{nPoisson}).

We have rephrased our previous (hard) problem of computing the commutators (\ref{commu}) of 
the tensor operators  (\ref{auarop2}) in terms of (more easy) Moyal brackets of 
functions on the coalgebra $u(N_+,N_-)^*$ [up to quotients by the ideals ${\cal 
I}$ generated by ``null-type" polynomials like (\ref{reord})]. 
Nevertheless, Moyal bracket captures the 
essence of more general deformations, which may include central extensions like  
\begin{eqnarray}
\left[\hat{L}^I_m, \hat{L}^J_n\right]=&&
\hbar\Lambda^{\alpha\beta}(J_\alpha m_\beta-
I_\alpha n_\beta)\hat{L}^{I+J-\delta_\alpha}_{m+n}+ {\rm O}(\hbar^3) \nn\\ &+&
\hbar^{(\sum_{\alpha=1}^N{I_\alpha+J_\alpha})}Q_I(m)\delta^{I,J}\delta_{m+n,0}
\mathbb I,
\end{eqnarray}
with central charges $Q_I(m)$ for all $U(N_+,N_-)$-spin $I$ currents 
$\hat{L}^I_m$. Note that, the structure of this central extension implies that the modes 
$\hat{L}^I_m$ and $\hat{L}^I_{-m}$ are conjugated, a fact inherited from the conjugation relation 
$\hat{X}_{\alpha\beta}^\dag=\hat{X}_{\beta\alpha}$ after (\ref{conjrel}) and the definition (\ref{auarop2}) of 
$\hat{L}^I_m$. An exhaustive study of this central extensions is in progress. 
Note that the diffeomorphism subalgebra $w^{(1)}_\infty(N_+,N_-)$ remains 
unaltered by Moyal deformations. 

\section{Towards a geometrical interpretation of ${\cL}_\infty(N_+,N_-)$\label{secoherente}}
In this Section we want to highlight the higher-spin structure of 
${\W}_\infty(N_+,N_-)$. To justify this view, we shall develop the representation theory of 
$U(N_+,N_-)$ (discrete series), calculating higher-spin representations, coherent states and deriving 
K\"ahler structures on flag manifolds, which are essential ingredients to 
define operator symbols. 

\subsection{Complex coordinates on flag manifolds\label{secoherente1}}
Although we shall restrict ourselves to the compact $SU(N)$ case in the 
following general discussion, most of the results are easily extrapolated 
to the non-compact $SU(N_+,N_-)$ case. Actually, we shall exemplify our 
construction with the (3+1)-dimensional conformal group $SU(2,2)=SO(4,2)$. 

In order to put coordinates on 
$G=SU(N)$, the ideal choice is the Bruhat decomposition 
\cite{Fulton} for the coset space (flag manifold)  
$\mathbb F=G/T$, where we denote $T=U(1)^{N-1}$ the maximal torus. 
We shall introduce a local complex parametrization of $\mathbb F$ by means 
of the isomorphism 
$G/T=G^{\mathbb C}/B$, where 
$G^{\mathbb C}\equiv SL(N,\mathbb C)$ is the complexification of $G$, and $B$ is 
the Borel subgroup of upper triangular matrices. In one direction, the 
element $[g]_T\in G/T$ is mapped to $[g]_B\in G^{\mathbb C}/B$. For 
example, for $G=SU(4)$ we have: 
\be
[g]_T=\bordermatrix{&\u1 &\u2 &\u3 &\u4 \cr &u_{11}&u_{12}&u_{13} &u_{14} 
\cr &u_{21}&u_{22}&u_{23}&u_{24} \cr &u_{31}&u_{32}&u_{33}&u_{34} \cr 
&u_{41}&u_{42}&u_{43}&u_{44} } \longrightarrow [g]_B=\bordermatrix{&\z1 
&\z2 &\z3 &\z4 \cr & 1 & 0&0 &0 \cr & z_{21} & 1&0&0&\cr& 
z_{31}&z_{32}&1&0&\cr& z_{41}&z_{42}&z_{43}&1}\label{triang} \ee where 
\bea & &z_{21}=\frac{u_{21}}{u_{11}},\,z_{31}=\frac{u_{31}}{u_{11}},\, 
z_{41}=\frac{u_{41}}{u_{11}},\nn\\ 
&&z_{32}=\frac{u_{11}u_{32}-u_{12}u_{31}}{u_{11}u_{22}-u_{12}u_{21}},\, 
z_{42}=\frac{u_{11}u_{42}-u_{12}u_{41}}{u_{11}u_{22}-u_{12}u_{21}}, 
\label{su22coord}\\ &&z_{43}=\frac{u_{13}(u_{21}u_{42}-u_{22}u_{41})- 
u_{23}(u_{11}u_{42}-u_{12}u_{41}) 
+u_{43}(u_{11}u_{22}-u_{12}u_{21})}{u_{13}(u_{21}u_{32}- 
u_{22}u_{31})-u_{23}(u_{11}u_{32}-u_{12}u_{31}) 
+u_{33}(u_{11}u_{22}-u_{12}u_{21})},\nn \eea provides a complex 
coordinatization $\{z_{\alpha\beta}, \alpha>\beta=1,2,3\}$ of nearly all 
of the 6-dimensional flag manifold 
$\mathbb F_{3}=SU(4)/U(1)^3$, missing only a lower-dimensional subspace; indeed, these coordinates are defined where the 
denominators are non-zero. In general, each flag $\mathbb F_{N-1}$ is 
covered by $N!$ patches, related to the elements of the Weyl group of $G$: 
the symmetric group $S_N$ of $N$ elements. A complete atlas of coordinate 
charts is obtained by moving this coordinate patch around by means of left 
multiplication with the Weyl group representatives (see e.g. 
\cite{Picken}). We shall restrict ourselves to the largest Bruhat cell 
(\ref{triang}). 

In the other direction, i.e. from 
$G^{\mathbb C}/B$ to $G/T$, one uses the Iwasawa decomposition: 
any element $g^c\in G^{\mathbb C}$ may be factorized as $g^c=gb, g\in G, 
b\in B$ in a unique fashion, up to torus elements $t\in T$ (the Cartan 
subgroup of diagonal matrices $t={\rm diag}(t_1,t_2/t_1,t_3/t_2,\dots, 
1/t_{N-1})$), which coordinates $t_\alpha$ can be calculated as the 
arguments $t_\alpha=(\Delta_\alpha(g)/\bar{\Delta}_\alpha(g))^{1/2}$ of 
the 
$\alpha$-upper principal minors $\Delta_\alpha$ of $g\in G$. For example, for 
$SU(4)$ we have:
 \bea &&t_1=\left(\frac{u_{11}}{\bar{u}_{11}}\right)^{1/2},\, 
t_2=\left(\frac{u_{11}u_{22}-u_{12}u_{21}}{\bar{u}_{11}\bar{u}_{22}- 
\bar{u}_{12}\bar{u}_{21}}\right)^{1/2},\\ 
&&t_3=\left(\frac{u_{13}(u_{21}u_{32}- 
u_{22}u_{31})-u_{23}(u_{11}u_{32}-u_{12}u_{31}) 
+u_{33}(u_{11}u_{22}-u_{12}u_{21})}{\bar{u}_{13}(\bar{u}_{21}\bar{u}_{32}- 
\bar{u}_{22}\bar{u}_{31})-\bar{u}_{23}(\bar{u}_{11}\bar{u}_{32}- 
\bar{u}_{12}\bar{u}_{31}) +\bar{u}_{33}(\bar{u}_{11}\bar{u}_{22} 
-\bar{u}_{12}\bar{u}_{21})}\right)^{1/2}.\nn \eea The Iwasawa 
decomposition in this case may be proved by means of the Gram-Schmidt 
ortonormalization process: regard any $g^c=[g]_B\in G^{\mathbb C}$ [like 
the one in (\ref{triang})] as a juxtaposition of $N$ column vectors 
$(\z1,\z2,\dots,{z}_N)$. Then one obtains orthogonal vectors $\{\w\alpha\}$ in the usual way:
\be
v_\alpha'=\left(\z\alpha- 
\frac{(\z\alpha,v_{\alpha-1})}{(v_{\alpha-1}',v_{\alpha-1}')} 
v_{\alpha-1}'-\dots -\frac{(\z\alpha,\w1)}{(\w1',\w1')}\w1'\right),\;\; 
v_\alpha=\frac{v_\alpha'}{ 
\left(\Lambda^{\alpha\alpha}(v_\alpha',v_\alpha')\right)^{1/2}},\label{g-s} 
\ee (not sum on $\alpha$) where $(\z\alpha,v_\beta)\equiv 
\bar{z}_{\alpha\mu}\Lambda^{\mu\nu}v_{\beta\nu}$ denotes a scalar product 
with metric $\Lambda$. At this point, it should be noted that the previous 
procedure can be straightforwardly extended to the non-compact case 
$G=SU(N_+,N_-)$ just by considering the indefinite metric $\Lambda={\rm 
diag}(1,\stackrel{N_+}{\dots},1,-1,\stackrel{N_-}{\dots},-1)$. Using a 
relativistic notation, we may say that the vectors $v_{1},\dots, v_{N_+}$ 
are ``space-like'' [that is, 
$(v_\alpha,v_\beta)=1$] whereas $v_{N_++1},\dots,  
v_{N}$ are ``time-like'' [i.e, 
$(v_\alpha,v_\beta)=-1$]; this ensures that 
$v\Lambda v^\dag=\Lambda$. For 
example, for $SU(2,2)$, the explicit expression of (\ref{g-s}) proves to 
be: \bea \w1&=&{|\Delta_1|}\left(\ba{c} 1 
\\ z_{21} \\ z_{31} \\ z_{41}\ea\right),\,\, 
\w2={|\Delta_1||\Delta_2|}\left(\ba{c} -{\bar{z}_{21}}+z_{32} 
{\bar{z}_{31}}+z_{42} {\bar{z}_{41}} \\ 1+z_{32} z_{21} 
{\bar{z}_{31}}-z_{31} {\bar{z}_{31}}+z_{42} z_{21} {\bar{z}_{41}}-z_{41} 
{\bar{z}_{41}} \\ z_{32}+z_{32} z_{21} {\bar{z}_{21}}- {\bar{z}_{21}} 
z_{31}+z_{42} z_{31} {\bar{z}_{41}}- z_{32} z_{41} {\bar{z}_{41}} \\ 
z_{42}+z_{42} z_{21} {\bar{z}_{21}}-z_{42} z_{31} {\bar{z}_{31}}- 
{\bar{z}_{21}} z_{41}+z_{32} {\bar{z}_{31}} z_{41}\ea \right) \nn\\ 
\w3&=&{|\Delta_2||\Delta_3|}\left(\ba{c} \left[ -{\bar{z}_{32}} 
{\bar{z}_{21}}- {\bar{z}_{42}} z_{43} {\bar{z}_{21}}+{\bar{z}_{31}}-z_{42} 
{\bar{z}_{42}} {\bar{z}_{31}}\right.\\ \left.+z_{32} {\bar{z}_{42}} z_{43} 
{\bar{z}_{31}}+{\bar{z}_{32}} z_{42} {\bar{z}_{41}}+z_{43} 
{\bar{z}_{41}}-z_{32} {\bar{z}_{32}} z_{43} {\bar{z}_{41}}\right] \\ 
\left[{\bar{z}_{32}}+{\bar{z}_{42}} z_{43}-z_{42} {\bar{z}_{42}} z_{21} 
{\bar{z}_{31}}+z_{32} {\bar{z}_{42}} z_{43}z_{21} 
{\bar{z}_{31}}-{\bar{z}_{42}} z_{43} z_{31} {\bar{z}_{31}}+ {\bar{z}_{42}} 
{\bar{z}_{31}} z_{41}\right.\\ \left.+{\bar{z}_{32}}z_{42} z_{21} 
{\bar{z}_{41}}- z_{32} {\bar{z}_{32}} z_{43} z_{21} {\bar{z}_{41}}+ 
{\bar{z}_{32}} z_{43} z_{31} {\bar{z}_{41}}- {\bar{z}_{32}} z_{41} 
{\bar{z}_{41}}\right] \\ \left[1-z_{42} 
{\bar{z}_{42}}+z_{32}{\bar{z}_{42}} z_{43}- z_{42} {\bar{z}_{42}} z_{21} 
{\bar{z}_{21}}+ z_{32} {\bar{z}_{42}} z_{43} z_{21} {\bar{z}_{21}}- 
{\bar{z}_{42}}z_{43} {\bar{z}_{21}} z_{31}\right.\\ \left.+ {\bar{z}_{42}} 
{\bar{z}_{21}} z_{41}+z_{42} z_{21} 
 {\bar{z}_{41}}-z_{32} z_{43} z_{21} {\bar{z}_{41}}+z_{43} z_{31}
{\bar{z}_{41}}-z_{41} {\bar{z}_{41}} \right] 
\\ \left[{\bar{z}_{32}} z_{42}+z_{43}-z_{32} {\bar{z}_{32}} z_{43}+
{\bar{z}_{32}} z_{42} z_{21} {\bar{z}_{21}}-z_{32} {\bar{z}_{32}} z_{43} 
z_{21} 
 {\bar{z}_{21}}+{\bar{z}_{32}} z_{43} {\bar{z}_{21}} z_{31}\right.\\
\left.-z_{42} z_{21}{\bar{z}_{31}}+z_{32} z_{43} z_{21} {\bar{z}_{31}} 
-z_{43} z_{31} {\bar{z}_{31}}-{\bar{z}_{32}} {\bar{z}_{21}} 
z_{41}+{\bar{z}_{31}}z_{41}\right] \ea \right) \nn\\ 
\w4&=&{|\Delta_3|}\left(\ba{c}-{\bar{z}_{42}} 
{\bar{z}_{21}}+{\bar{z}_{32}} {\bar{z}_{43}} {\bar{z}_{21}}-{\bar{z}_{43}} 
{\bar{z}_{31}}+{\bar{z}_{41}}\\ {\bar{z}_{42}}-{\bar{z}_{32}} 
{\bar{z}_{43}}\\ -{\bar{z}_{43}}\\ 1\ea\right) \eea where 
\begin{eqnarray}
|\Delta_1(z,\bar{z})|&=&\frac{1}{\sqrt{1+|z_{21}|^2-|z_{31}|^2 
-|z_{41}|^2}} \label{lengths}\\ 
|\Delta_2(z,\bar{z})|&=&\frac{1}{\sqrt{1+|z_{32}z_{41}-z_{42}z_{31}|^2-|z_{32}|^2- 
|z_{42}|^2 -|z_{32}z_{21}-z_{31}|^2-|z_{42}z_{21}-z_{41}|^2}} \nn\\ 
|\Delta_3(z,\bar{z})|&=&\frac{1}{\sqrt{1+|z_{43}|^2-|z_{42}-z_{43}z_{32}|^2-|z_{41}+ 
z_{43}z_{32}z_{21}-z_{42}z_{21}-z_{43}z_{31}|^2}}\nn 
\end{eqnarray}
are the moduli of the $\alpha=1,2,3$ upper principal minors 
$\Delta_\alpha(g)$ of $g\in G$. These ``characteristic lengths" will 
play a central role in what follows. 

 Any (peudo-) unitary  matrix $g\in G$ in the present patch (which contains the 
identity element $z=0=\bar{z},t=1$) can be written in minimal coordinates
$g=(z_{\alpha\beta}, \bar{z}_{\alpha\beta},
t_\beta),\,\alpha>\beta=1,\dots,N-1$, as the product $g=vt$ of an element 
$v$ of the base (flag) $\mathbb F$ times an element $t$ of the fibre 
$T=U(1)^{N-1}$. 

Once we have the expression of a general $G$ group element 
$g=(g^1,\dots,g^{N^2-1})$ in terms of the minimal coordinates 
$g=(z_{\alpha\beta},\bar{z}_{\alpha\beta},t_\beta), \alpha>\beta=1,\dots, N-1$,
we can easily write the group law $g''=g'\bullet g$ and compute the left- 
and right-invariant vector fields \be\left.\begin{array}{ll} 
X^L_j(g)\equiv {\cal L}_j^k(g) \frac{\partial}{\partial g^k}, & {\cal 
L}_j^k(g)= \left.\frac{\partial (g\bullet g')^k}{\partial 
g'^j}\right|_{g'=e}\\ X^R_j(g)\equiv {\cal R}_j^k(g) 
\frac{\partial}{\partial g^k}, & {\cal R}_j^k(g)= \left.\frac{\partial 
(g'\bullet g)^k}{\partial g'^j}\right|_{g'=e}\end{array}\right\} 
j,k=1,\dots,N^2-1={\rm dim}(G).\label{lrivf}\ee The algebraic 
correspondence between right-invariant vector fields and the step 
operators (\ref{pun}), with commutation relations (\ref{conjrel}), is: 
\be
X^{R}_{z_{\alpha\beta}}\to \hat{X}_{\alpha\beta},\;\; 
X^{R}_{\bar{z}_{\alpha\beta}}\to \hat{X}_{\beta\alpha},\;\; 
X^{R}_{t_{\beta}}\to \hat{X}_{\beta\beta}- 
\hat{X}_{\beta+1,\beta+1},\;\;\; \alpha>\beta=1,\dots,N-1. \ee 

\subsection{Higher-spin representations, coherent states and K\"ahler structures on flag manifolds\label{secoherente2}}
In this Section we shall compute the unitary irreducible representations 
of $G$ and we shall construct coherent states and geometric structures 
attached to them. Let us start by considering the (finite) left regular 
representation 
$[L_{g}\Psi](g')=\Psi({g}^{-1}\bullet g')$ of the group 
$G$ on complex functions $\Psi$ on $G$ [remember Eq. (\ref{analcomp})]. 
This representation is highly reducible. The reduction can be achieved 
through a complete set of finite right restrictions or ``polarization 
equations'' (in the language of geometric quantization \cite{gq}): 
\begin{equation}
[R_g\Psi](g')=\Psi(g'\bullet g)=D^c(g)\Psi(g'),\qquad \forall g\in 
P,\forall g'\in G,\label{poleq1} 
\end{equation}
which impose that $\Psi$ must transform according to a given (Abelian) 
representation $D^c$ (with index $c$) of a certain maximal proper subgroup 
$P\subset G$ (``polarization subgroup"). The Lie algebra 
$\cal P$ of $P$ is called a ``first-order polarization'', which formal 
definition could be stated as 
\begin{defn}
A first-order polarization is a proper subalgebra ${\cal P}$ of the Lie 
algebra ${\cal G}$ of $G$, realized in terms of left-invariant vector 
fields $X^L$ [the infinitesimal generators of finite right translations 
(\ref{poleq1})]. It must satisfy a maximality condition in order to define 
an irreducible representation of $G$. 
\end{defn}
Hence, at the Lie algebra level, the polarization equations (\ref{poleq1}) 
acquire the form of a system of non-homogeneous first-order partial 
differential equations:\footnote{This procedure for obtaining irreducible 
representations resembles Mackey's induction method, except for the fact 
that it can be extended to ``higher-order polarizations'': subalgebras 
${\cal P}^{HO}$ of the (left) universal enveloping algebra ${\cal U(G)}$ 
which also satisfy a maximality condition in order to define an 
irreducible representation (see e.g. \cite{higher-order} for more 
details)} 
\be
X^L_j\Psi=c_j\Psi(g)\,,\qquad \forall X^L_j\in {\cal P} \label{poleq2} \ee 
where $c$ denotes a one-dimensional representation (character) of the 
polarization subalgebra ${\cal P}$, 
$c(X^L_i)=c_i\,,\forall X^L_i\in {\cal P}$. That is,  $c$ is the 
infinitesimal character associated to $D^c$ in (\ref{poleq1}). Notice that 
since the representation is one-dimensional, the character $c$ vanishes on 
the derived subalgebra $[{\cal P},{\cal P}]$ of ${\cal P}$, i.e.  
$c([X^L_i,X^L_j])=0\,,\forall X^L_i,X^L_j\in {\cal P}$.  This means that the character $c$ can be 
non-zero only on the quotient 
${\cal P}/[{\cal P},{\cal P}]$, which is an Abelian algebra. 
For our case, the first-order polarization subalgebra 
$\cal P$ is generated by 
the following $(N-1+N(N-1)/2)$ left-invariant vector fields 
\[{\cal P}=\langle  X^L_{t_\beta}, X^L_{z_{\alpha\beta}}, \alpha>\beta=1,\dots,N-1\rangle. \]
Then, the quotient ${\cal P}/[{\cal P},{\cal P}]$ coincides here with the 
Abelian Cartan subalgebra ${\cal T}=u(1)^{N-1}$. Therefore, denoting by 
$c( X^L_{t_\beta})\equiv -2S_\beta, \forall X^L_{t_\beta}\in {\cal T}$ the 
non-zero characters or ``$G$-spin labels", the solution to the 
polarization equations (\ref{poleq2}), 
\be
\left.\ba{l} X^L_{t_\beta}\Psi=-2S_\beta\Psi \\ 
X^L_{z_{\alpha\beta}}\Psi=0,\ea\right\}\Rightarrow 
\Psi^{S}(g)=W_{S}(g)\Phi(\bar{z}), \label{antiholf} \ee can be arranged as 
the product of a highest-weight vector $W_S$ (``vacuum''), which is a 
particular solution of $X^L_{t_\beta}\Psi=-2S_\beta\Psi$ and can be 
written as a product of upper principal minors: \be W_{S}(g)\equiv 
\prod_{\beta=1}^{N-1} \left(\bar{\Delta}_\beta(g)\right)^{2S_\beta},\ee 
times an anti-holomorphic function $\Phi(\bar{z})$, which can be written 
as an analytic power series, with complex coefficients $a^{S}_{m}$, on its 
arguments $\bar{z}_{\alpha\beta}$, \be\Phi(\bar{z})\equiv \sum_{m} 
a^{S}_{m}\prod_{\alpha>\beta} 
(\bar{z}_{\alpha\beta})^{m_{\beta\alpha}}\label{vacsu22}. \ee The index 
$m$ denotes an integral upper-triangular $N\times N$ matrix [see  
(\ref{uppert})]. The range of the entries $m_{\alpha\beta}, 
\alpha<\beta=2,\dots,N$ depends on the set of $G$-spin indices  
$\{S_\beta\}_{\beta=1}^{N-1}$, which label particular $G$-spin $S$ 
irreducible representations of $G$ on the Hilbert space 
${\cal H}_S(G)$ of polarized wave functions (\ref{antiholf}). 

The sign of the $SU(N_+,N_-)$-spin indices $S_\beta$ depends on the 
(non-)compact character of the corresponding simple roots: the ones whose 
generators $X_{\alpha\beta}$ fulfil $\beta=\alpha+1$. With this notation, 
all the roots $(\alpha\beta)$ are of compact type except for 
$(\alpha\beta)=(N_+,N_++1)$. This fact implies that $S_\beta\in\mathbb Z^+/2$
except for $S_{N_+}\in\mathbb Z^-/2$. Indeed, with this choice of sign we 
guarantee: a) the \emph{finiteness} of the scalar product 
$\langle\Phi|\Psi\rangle\equiv\int_G d^Lg\,\bar{\Phi}(g) \Psi(g)$, which 
Haar measure has the form:
\be
d^Lg={\prod_{\beta=1}^{N-1}|\Delta_\beta(z,\bar{z})|^4} 
\bigwedge_{\beta=1}^{N-1}t_\beta^{-1}dt_\beta \bigwedge_{\alpha>\beta}d 
z_{\alpha\beta}\wedge d \bar{z}_{\alpha\beta}, \label{Haar}\ee [where we 
have used that 
$\det({\cal L}_j^k(g))^{-1}={\prod_{\beta=1}^{N-1}|\Delta_\beta(z,\bar{z})|^4t_\beta^{-1}}$] 
and b) the \emph{unitarity} of the representation 
$[L_{g'}\Psi](g)=\Psi({g'}^{-1}\bullet g)$ of $G$. We can still keep track 
of the extra $U(1)$ quantum number 
$S_N$ that differentiates $U(N)\simeq (SU(N)\times U(1))/\mathbb Z_N$ from 
$SU(N)$ representations. The $U(N)$ wave functions 
$\tilde{\Psi}^{I}$ depend on an extra $U(1)$-factor 
$(t_N)^{-2S_N}, t_N\in U(1)$ in the vacuum wave function $W_{S}$
in (\ref{vacsu22}), where the relation between the $U(N)$-spin labels 
$I=(I_1,\dots,I_N)$ of Eq. (\ref{auarop2}) and the 
$SU(N)\times U(1)$-spin labels $S=(S_1,\dots,S_N)$ is: 
$S_\beta=I_\beta-I_{\beta+1}, \beta=1,\dots,N-1$ and
$S_N=\sum_{\alpha=1}^NI_\alpha$ [the Casimir $C_1$ (trace) eigenvalue].

The basic wave functions 
$\Psi^{S}_m(g)\equiv
W_{S}(g)\prod_{\alpha>\beta}(\bar{z}_{\alpha\beta})^{m_{\beta\alpha}}$ of 
${\cal H}_S(G)$ are eigenfunctions of the right-invariant differential operators 
$X^R_{t_\beta}$ (Cartan generators):
\be
X^R_{t_\beta}\Psi^{S}_m=(2S_\beta+m_\beta-m_{\beta+1})\Psi^{S}_m, 
\label{cartanact} \ee where $m_\beta$ is defined in (\ref{vecupper}); 
notice that the eigenvalue 
$(2S_\beta+m_\beta-m_{\beta+1})$ of $X^R_{t_\beta}$ can also be written as 
$2(\Gamma^0_\beta(S,m)-\Gamma^0_{\beta+1}(S,m))$, where
$\Gamma^0_\beta(S,m)$ is one of the characteristic
factors (\ref{factors}) that appears in the power expansion of the 
structure constants (\ref{hosc}) of the algebra (\ref{qnPoisson}). The 
lowering operators $Z_{\alpha\beta}\equiv X^R_{\bar{z}_{\alpha\beta}}$ 
annihilate the vacuum vector $\Psi^S_0=W_S$. The rest of vectors 
$L^S_m(g)$ of the Hilbert space 
${\cal H}_S(G)$ can be obtained through the orbit of the vacuum under the action of 
rising operators $Z^\dag_{\alpha\beta}\equiv X^R_{{z}_{\alpha\beta}}$: \be 
L^S_m(g)\equiv\prod_{\alpha>\beta} 
(Z^\dag_{\alpha\beta})^{m_{\beta\alpha}}W_S(g),\;m_{\alpha\beta}\in 
\mathbb N.\ee Notice that the way of labelling the enveloping algebra 
operators (\ref{auarop2}) and base vectors 
$L^S_m$ in the carrier space ${\cal H}_S(G)$ of irreducible representations of $G$ coincides: 
the upper $G$-spin index $S$ is an integral vector and the lower index 
(``third component") 
$m$ is an integral upper-triangular matrix). Negative modes 
$\hat{L}^I_{-|m|}$ in (\ref{auarop2}) would correspond to the complex conjugate (holomorphic) vectors 
$L^S_{-m}\equiv\bar{L}^S_{m}$. We shall give later on Equation (\ref{Wtransf}) the explicit expression of the orbit 
${\cal F}_0=\{L_g\Psi^S_0, g\in G\}$ of the vacuum vector $\Psi^S_0=W_S$ under the finite left action of the group $G$. 

Denote $\langle{S g}|{\Psi}\rangle\equiv \Psi^S(g)$ and 
$\langle{\Psi}|{S g}\rangle\equiv \bar{\Psi}^S(g)$. The coherent state overlap or ``reproducing kernel''  
$\Delta_{S}(g,g')\equiv \scprod{S g}{S g'}$ can be calculated by inserting the
resolution of unity 
\be
1=\sum_m \ket{\chi_m}\bra{\chi_m} \ee given by an orthonormal basis 
$\{|\chi_m\rangle\}$ of ${\cal H}_S(G)$. The explicit expression of this overlap in terms of 
upper-minors $\Delta_\beta, \beta=1,\dots,N-1$, of $g=(t,z,\bar{z})\in G$ 
turns out to be: 
\be
\Delta^{S}(g,g')=\sum_m{ 
\chi^S_m(g)\bar{\chi}^{S}_m(g')}=\prod_{\beta=1}^{N-1} 
\frac{(\bar{t}_\beta|\Delta_\beta(z,\bar{z})|)^{2S_\beta}({t}'_\beta 
|\Delta_\beta(z',\bar{z}')|)^{2S_\beta}}{|\Delta_\beta(z',\bar{z})|^{4S_\beta}}.\label{repker}\ee 
This reproducing kernel satisfies the integral equation of a projector 
operator 
 \be
 \Delta^{S}(g,g'')=\int \Delta^{S}(g,g')\Delta^{S}(g',g'')d^Lg'
 \ee
and the propagator equation: 
\be
\Psi^{S}(g)=\int_G{d^Lg'\,\Delta^{S}(g,g')\Psi^{S}(g')}, \ee where we have 
used the resolution of unity 
\be
1=\int_G{d^Lg\,\ket{S g}\bra{S g}}. \ee 

Given a vector $\gamma\in {\cal H}_S(G)$ (for example the vacuum 
$W_S(g)\equiv \langle Sg|0\rangle$) the set of vectors in the orbit of $\gamma$ under $G$, 
${\cal F}_\gamma=\{\gamma_g=L_{g}\gamma, g\in 
G\}$, is called a family of covariant CS. We know from (\ref{cartanact}) 
that the Cartan (\emph{isotropy}) subgroup $T=U(1)^{N-1}$ stabilizes the 
vacuum vector 
$\gamma=W_S$ up to  multiplicative phase factors 
$t^{2S_\beta}_\beta$ (characters of $T$). Actually, the explicit expression of the family 
${\cal F}_\gamma$ of CS for $\gamma=W_S$ turns out to be: 
 \begin{equation}
 [L_{g}W_S](g')=W_S(g^{-1}\bullet g')=W_S(g')e^{-\Theta_S(\bar{z}',g)}\prod_{\beta=1}^{N-1} 
 t_\beta^{2S_\beta},\label{Wtransf}
 \end{equation}
where we define 
\begin{equation}
\Theta_S(\bar{z}',g)\equiv-\sum_{\beta=1}^{N-1} 
2S_\beta\ln\frac{|\Delta_\beta(z,\bar{z})|}{|\Delta_\beta(z,\bar{z}')|^2} 
\end{equation}
an anti-holomorphic function of $\bar{z}'$ fulfilling cocycle properties 
(see bellow) and related to the so called ``multipliers'' (Radon-Nikodym 
derivative) in standard representation theory. 

Considering the flag manifold 
$\mathbb F=G/T$ and taking the Borel section 
$\sigma: \mathbb F\to G, \sigma(z,\bar{z})=(z,\bar{z},t=1)=g$ (which 
appears implicitly in the factorization $g=vt$) we may define another 
family of covariant CS as 
$\gamma_{\sigma(z,\bar{z})}=L_{\sigma(z,\bar{z})}\gamma$ (classes of CS 
modulo $T$), which are usually referred to as the Gilmore-Perelomov CS. 

It is also known in the literature that the flag manifold $\mathbb F$ is a 
K\"ahler manifold, with local complex coordinates 
$z_{\alpha\beta}, \bar{z}_{\alpha\beta}$ (\ref{su22coord}), an Hermitian Riemannian metric $\eta$ 
and a corresponding closed two-form (K\"ahler form) $\Omega$, 
\be
ds^2=\eta^{\alpha\beta,\mu\nu} dz_{\alpha\beta}d\bar{z}_{\mu\nu},\;\; 
\Omega=i \eta^{\alpha\beta,\mu\nu}dz_{\alpha\beta}\wedge 
d\bar{z}_{\mu\nu},\label{kahlerform} \ee which can be obtained from the 
K\"ahler potential 
\be
K_{S}(z,\bar{z})\equiv-\sum_{\beta=1}^{N-1} 4S_\beta\ln 
|\Delta_\beta(z,\bar{z})| \label{kahlerpot} \ee through the formula 
$\eta^{\alpha\beta,\mu\nu}=\frac{\partial}{\partial z_{\alpha\beta}}
\frac{\partial}{\partial \bar{z}_{\mu\nu}}K_{S}$. Notice that the K\"ahler 
potential $K_{S}$ essentially corresponds to the natural logarithm of the 
squared vacuum modulus 
$K_S(z,\bar{z})=-\ln|W_{S}(z,\bar{z},t)|^2$ in (\ref{vacsu22}). Actually, given the 
holomorphic action of $G$ on $\mathbb F$, \[(z',\bar{z}')\to 
g(z',\bar{z}')=\sigma^{-1}(g^{-1}\bullet (z',\bar{z}',1)),\, g\in G,\, 
(z',\bar{z}')\in \mathbb F,\] the transformation properties of $K_S$ are 
inherited from those of $W_S$ in (\ref{Wtransf}): 
\begin{equation}
K_S(gz,\overline{gz})=K_S(z,\bar{z})+\Theta_S(\bar{z},g)+\overline{\Theta_S(\bar{z},g)}.
\end{equation} 
The function $\Theta_S$ verifies the cocycle condition 
$\Theta_S(\bar{z},g'\bullet g)=\Theta_S(\overline{gz},g')+\Theta_S(\bar{z},g)$, 
which results from the group property $g'(gz)=(g'\bullet g)z$. 

\subsection{Operator symbols on flag manifolds}

Let us consider the finite left translation 
$[L_{g'}\Psi^S](g)\equiv\Psi^S(g'^{-1}\bullet g)$ as a linear operator in ${\cal H}_S(G)$. The symbol 
[remember the definition (\ref{nondigsymb})] $L_{g'}^S(g,h),\, g,g',h\in 
G$ of the operator $L_{g'}$ representing the group element $g'\in G$ in 
${\cal H}_S(G)$ can be written in terms of the reproducing kernel (\ref{repker}) as: 
\begin{equation}
L_{g'}^S(g,h)=\langle Sg|L_{g'}|Sh\rangle=\Delta^S(g'^{-1}g,h). 
\end{equation}
Knowing that right-invariant vector fields $X^R$ [defined in 
(\ref{lrivf})] are the infinitesimal generators of finite left 
translations $L_g$, one can easily compute the symbols 
$X_j^S(g,h)$ of the Lie-algebra ${\cal G}$ generators $\hat{X}_j$ as:
\begin{equation}
X_j^S(g,h)\equiv\langle Sg|\hat{X}_j|Sh\rangle=X^R_j(g)\Delta^S(g,h)= 
{\cal R}_j^k(g)\frac{\partial}{\partial g^k}\Delta^S(g,h)\label{symbolsG}. 
\end{equation}
From a quantum-mechanical perspective, the points $g\in G$ do not label 
distinct states $|g\rangle=L_g|0\rangle$ because of the inherent phase 
freedom in quantum mechanics. Rather, the corresponding quantum state 
depends on its equivalence class $(z,\bar{z})=gT$ modulo $T$. Let us 
consider then the new action of $G$ on the anti-holomorphic part 
$\Phi(\bar{z})$ of $\Psi^S(g)$ in (\ref{antiholf}). Since the vacuum 
$W_S$ is a fixed common factor of 
all the wave functions $\Psi^S=W_S\Phi$ in (\ref{antiholf}), we can factor 
it out and consider the restricted action ${L}_g^S\equiv W_S^{-1}L_gW_S$ 
on the arbitrary anti-holomorphic part $\Phi(\bar{z})$, thus resulting in: 
\begin{equation}
[{L}_g^S \Phi](\bar{z}')=e^{-\Theta_S(\bar{z}',g)}\Phi(g^{-1}\bar{z}'), \, 
g=(z,\bar{z},t)\in G 
\end{equation}
(modulo $T$). The infinitesimal generators $X^S_j$ of this new restricted 
action can be written as: 
\begin{equation} 
X^S_j=\nabla_j-\theta_j^S(\bar{z}), 
\end{equation} 
where  
\[\nabla_j\equiv X^R_j(g)(g\bar{z})_{\alpha\beta}|_{g=e}\frac{\partial}{\partial 
\bar{z}_{\alpha\beta}},\;\;\theta_j^S(\bar{z})\equiv 
X^R_j(g)\Theta_S(\bar{z},g)|_{g=e}.\] Denoting now $\langle 
\bar{z}|\Phi\rangle\equiv \Phi(\bar{z}), \langle \Phi|\bar{z}\rangle\equiv 
\bar{\Phi}(\bar{z})$ and $L^S_g(\bar{z},z')\equiv \langle 
\bar{z}|L^S_g|\bar{z}'\rangle$, the restriction of the symbols 
(\ref{symbolsG}) to the flag manifold $\mathbb F$ can be written in terms 
of the K\"ahler potential $K_S$ and the cocycle 
$\Theta_S$ as follows: 
\begin{equation}
X^S_j(\bar{z},z')=X^R_j(g)L^S_g(\bar{z},z')|_{g=e}=\nabla_jK_S(\bar{z},z')-\theta_j^S(\bar{z}).
\end{equation} 
The diagonal part $X^S_j(\bar{z},z)$ are called \emph{equivariant momentum 
maps}. Using Lie equations for 
$\nabla_j$ and differential properties of the cocycle $\theta_j^S$, one 
can prove that momentum maps implement a realization of the Lie algebra 
$\cal G$ of $G$ in terms of Poisson brackets (\ref{poiscoad}). 

The correspondence between commutator (\ref{conjrel}) and Poisson bracket 
(\ref{poiscoad}) does not hold in general for arbitrary elements like 
(\ref{auarop2}) in the universal enveloping algebra ${\cal U}({\cal G})$. 
As we stated in Equation (\ref{comsymb}), the star commutator of symbols 
admits a power series expansion in the $G$-spin parameters 
$S_\beta$ (being the Poisson bracket the leading term), so that star commutators 
converge to Poisson brackets for large quantum numbers $S\to \infty$.  

We believe that higher-order terms in the Moyal commutators 
(\ref{qnPoisson}) give a ``taste'' of these higher order corrections to 
the Poisson bracket in the star commutator (\ref{comsymb}) of symbols, 
which actual expression seems hard to compute. 

\section{Field Models on Flag Manifolds\label{fieldflag}}
Before finishing, we would like to propose some interesting applications 
like diffeomorphism invariant field models, based on Yang-Mills theories, 
and non-linear sigma models on flag manifolds. 
\subsection{Volume-preserving diffeomorphisms and higher-extended objects 
\label{generalization}} 

We showed in Sec. \ref{su2largeN} that the low-energy limit of
the $SU(\infty)$ Yang-Mills action (\ref{suinftyaction}), 
described by space-constant (vacuum configurations) $SU(\infty)$ vector potentials
$X_\mu(\tau;\vartheta,\varphi)\equiv A_\mu(\tau,\vec{0};\vartheta,\varphi)$,
turns out to reproduce the dynamics of the relativistic spherical membrane 
$\mathbb F_1=S^2$ . This view can be straightforwardly extended to arbitrary 
flag manifolds $\mathbb F_{N-1}=SU(N)/U(1)^{N-1}$ just replacing the 
Poisson bracket on the sphere (\ref{spherepb}) by (\ref{poiscoad}). 
Actually, as it is done for ${\rm SDiff}(S^2)$ gauge invariant Yang-Mills 
theories in (\ref{suinftyaction}), an action functional for a 
${\rm SDiff}(\mathbb F_{N-1})$ gauge invariant Yang-Mills theory in four 
dimensions could be written as: 

\bea S&=&\int {\rm d}^4x \langle F_{\nu\gamma}| 
F^{\nu\gamma}\rangle\,,\nn\\ F_{\nu\gamma}&=&\partial_\nu 
A_\gamma-\partial_\gamma A_\nu + \left\{A_\nu,A_\gamma\right\}_P\,,\\ 
A_\nu(x;\bar{z},z)&=&\sum_{\{S,m\}}A_{\nu S}^m(x) 
L^S_m(\bar{z},z)\,,\;\;\nu,\gamma=1,\dots,4\,,\nn \eea where now  
$\langle \cdot|\cdot\rangle$ denotes the scalar product between tensor operator symbols $L^S_m$ on $\mathbb F_{N-1}$, with 
integration measure (\ref{intmeasleaf}), which explicit expression is 
straightforwardly obtained from the left-invariant Haar measure 
(\ref{Haar}) on the whole group $G$ after inner derivation 
$i_{X}$ by left-invariant generators of toral (Cartan $T$) elements:
\[d\mu(z,\bar{z})=\prod_{\beta=1}^{N-1}i_{X^L_{t_\beta}}d^Lg={\prod_{\beta=1}^{N-1}|\Delta_\beta(z,\bar{z})|^4} 
\bigwedge_{\alpha>\beta}d z_{\alpha\beta}\wedge d \bar{z}_{\alpha\beta}.\] 
Hence, all (infinite) higher-$G$-spin $S$ vector fields $A_{\nu S}^m(x)$ 
on 
$\mathbb{R}^4$ are combined into a single field 
$A_\nu(x;z,\bar{z})$ on the extended manifold $\mathbb{R}^4\times \mathbb F_{N-1}$; that is,  
$A_{\nu S}^m(x)$ can be considered as a particular ``vibration mode of the $N(N-1)$-brane" 
$\mathbb F_{N-1}$.

In the same way, a 2+1-dimensional Chern-Simons ${\rm SDiff}(\mathbb 
F_{N-1})$-invariant gauge theory can be formulated with action: 
\be
S=\int_{\mathbb{R}^3\times \mathbb F_{N-1}} (A\wedge d 
A+\frac{1}{3}\{A,A\}\wedge A),\;\;A=A_\mu dx^\mu, \ee and equations of 
motion: $F=0$. 
\subsection{Nonlinear sigma models on flag manifolds}
Let us consider a matrix $v\in SU(N)/T$ (as a gauge group, i.e. as a map 
$v:\mathbb R^D\to SU(N)$), which is a juxtaposition 
$v=({v}_1,\dots,{v}_N)$ of the $N$ orthonormal vectors 
${v}_\alpha$ in (\ref{g-s}). The Maurer-Cartan form can be decomposed in 
diagonal and off-diagonal parts
\begin{equation}
v^{-1}dv=v^\dag dv=\left(\ba{c} \bar{v}_1^t\\ \vdots \\ 
\bar{v}_N^t\ea\right) \left(\ba{ccc} dv_1, &\cdots, &dv_N 
\ea\right)=\sum_{\alpha=1}^N \bar{v}^t_\alpha d v_\alpha 
\hat{X}_{\alpha\alpha}+\sum_{\alpha\not=\beta} \bar{v}^t_\alpha d v_\beta 
\hat{X}_{\alpha\beta}, 
\end{equation}
where $\hat{X}_{\alpha\beta}$ are the step operators (\ref{bosoprea}). The 
Lagrangian density for the non-linear sigma model (SM) on the coset (flag) 
$G/T$ 
\begin{equation}
{\cal L}_{SM}=\frac{\kappa}{8}{\rm tr}_{G/T}(v^{-1}\partial_\mu v 
v^{-1}\partial_\mu v) 
\end{equation}
is written in terms of the off-diagonal parts as
\begin{equation}
{\cal L}_{SM}=\frac{\kappa}{2}\sum_{\alpha<\beta}(v_\alpha,\partial_\mu 
v_\beta)^2.\label{nlsm} 
\end{equation}
The usual Lagrangian for the complex projective space $\mathbb 
CP^{N-1}=SU(N)/(SU(N-1)\times U(1))$ 
\begin{equation} 
{\cal L}_{\mathbb CP^{N-1}}=\frac{\kappa}{2} 
\eta_{\alpha\bar{\beta}}(\varphi)\frac{\partial \varphi^\alpha}{\partial 
x^\mu} \frac{\partial \bar{\varphi}^\beta}{\partial x_\mu} 
,\,\,\eta_{\alpha\bar{\beta}}(\varphi)\equiv 
\delta_{\alpha\beta}-{\varphi}_\alpha\bar{\varphi}_\beta,\,\,\varphi^\dag\cdot\varphi=1 
\label{cpnm} 
\end{equation}
can be also obtained as a particular case of (\ref{nlsm}) as follows. 
Unitary matrices $w$ on the coset $\mathbb CP^{N-1}$ are obtained from 
$[g]_B$ in (\ref{triang}) by considering the particular local complex 
parametrization where $z_{\alpha\beta}=0, \forall\beta\geq 2$. Let us 
consider a new basis 
$\{\hat{J}^k, k=1,\dots, N^2-1\}$ of traceless hermitian matrices for the 
Lie algebra $su(N)$, normalized as ${\rm 
tr}(\hat{J}^k\hat{J}^l)=\frac{1}{2}\delta_{kl}$. Let us use 
$|0\rangle$ for the Dirac notation for the vacuum vector 
$W_S(g)=\langle Sg|0\rangle$. In the fundamental representation (lowest $S$), and for the $\mathbb CP^{N-1}$ case, 
the vacuum is given by the 
 column $N$-vector \[|0\rangle=\left(\ba{cccc} 1,&  0, & \cdots, &  
0\ea\right)^t.\] If we define by \[q^k\equiv\langle 
0|w\hat{J}^kw^\dag|0\rangle=(w\hat{J}^kw^\dag)_{11}\] the vacuum 
expectation value of the conjugated Lie algebra element 
$w\hat{J}^kw^\dag$ under the adjoint action of the group $G$, then the restriction 
\[{\cal L}_{\mathbb CP^{N-1}}=\frac{\kappa}{2}\sum_{\beta=2}^N(w_1,\partial_\mu 
w_\beta)^2\] of (\ref{nlsm}) to $\mathbb CP^{N-1}$ could also be written 
as 
\[{\cal L}_{\mathbb CP^{N-1}}=\frac{\kappa}{2}\partial_\mu 
q\cdot\partial^\mu q,\] which coincides with (\ref{cpnm}) when we identify 
$\varphi_\alpha\equiv {w}_{\alpha 1}=z_{\alpha 1}|\Delta_1(z,\bar{z})|, \, 
z_{11}\equiv 1$. In particular, with this change of variable, one can see 
that the metric 
$\eta_{\alpha\bar{\beta}}$ in (\ref{cpnm}) coincides with $\eta^{\alpha 1;\beta 1}$ in 
(\ref{kahlerform}) for the restriction 
$K_{S_1}(z,\bar{z})\equiv -4S_1\ln 
|\Delta_1(z,\bar{z})|$ of the K\"ahler potential to $\mathbb CP^{N-1}$. 
\section*{Conclusions and outlook}
We provided a general view of, what we agreed to call, ``generalized 
${\cal W}_\infty$ symmetries", from various perspectives and approaches. 
We started discussing the structure of these new infinite-dimensional 
$\W$-like Lie algebras inside a group theoretical framework as algebras of 
$U(N_+,N_-)$ tensor operators. Inside this context, the (hard) problem of computing commutators 
of tensor operators has been rephrased in terms of (more easy) Moyal 
brackets of (polynomial) functions on the coalgebra $u(N_+,N_-)^*$, up to 
quotients by the ideals generated by null-type polynomials like 
(\ref{reord}). That is, we have intended to recover quantum commutators 
from quantum (Moyal) deformations of classical (oscillator) brackets. 
Moyal bracket captures the essence of the full (quantum) algebra, and 
makes use of the standard oscillator realization of the basic 
$u(N_+,N_-)$-Lie algebra generators. The resulting infinite-dimensional 
generalized 
$\W$-algebras can be seen as: 
\begin{enumerate}
\item infinite continuations of the finite-dimensional symmetries 
$u(N_+,N_-)$, or as
\item higher-$U(N_+,N_-)$-spin extensions of the 
diffeomorphism algebra diff$(N_+,N_-)$ of a $N$-dimensional manifold (e.g. 
a $N$-torus). 
\end{enumerate}
In order to justify the view of ${\W}_\infty(N_+,N_-)$ as a ``higher-spin 
algebra'' of $U(N_+,N_-)$, we have computed higher-spin representations of 
$U(N_+,N_-)$ (discrete series), we have given explicit expressions for coherent states and we 
have derived K\"ahler structures on 
flag manifolds, which are essential ingredients to define operator 
symbols. 

These infinite-dimensional Lie algebras potentially provide a new arena 
for integrable field models in higher dimensions, of which we have briefly 
mentioned gauge dynamics of higher-extended objects and reminded 
non-linear SM on flag manifolds. An exhaustive study of central extensions 
of ${\W}_\infty(N_+,N_-)$ should give us an important new ingredient 
regarding the constructions of unitary irreducible representations and 
invariant geometric action functionals, just as central extensions of 
standard $\W$ and Virasoro algebras encode essential information. This 
should be our next step.

\end{document}